\RequirePackage{fix-cm}
\documentclass[natbib]{svjour3}                     
\smartqed  
\usepackage{amsmath}
\usepackage{graphicx,psfrag,epsf}
\usepackage{enumerate}
\usepackage{url} 
\usepackage[T1]{fontenc}
\usepackage[latin9]{inputenc}
\usepackage{geometry}

\usepackage{amsthm}
\usepackage{amssymb}
\usepackage{graphicx}

\usepackage[linesnumbered,ruled,vlined]{algorithm2e}
\SetKwInput{KwInput}{Input}               	
\SetKwInput{KwOutput}{Output}			

\usepackage{expl3,xparse}
\usepackage{subcaption}
\theoremstyle{plain}
\newtheorem{thm}{\protect\theoremname}[section]
\theoremstyle{remark}
\newtheorem{rem}{\protect\remarkname}[section]

\begin{document}

\title{Sequential Monte Carlo with transformations\thanks{
    The authors gratefully acknowledge funding from BBSRC, the Wellcome Trust, the Royal Society and the Modernising Medical Microbiology group.}
}


\author{Richard G. Everitt \and Richard Culliford \and Felipe Medina-Aguayo \and Daniel J. Wilson
}


\institute{R. G. Everitt \and R. Culliford \and F. Medina-Aguayo \at
              Department of Mathematics and Statistics, University of Reading \\
              \email{richard.g.everitt@gmail.com}           
           \and
           D. J. Wilson \at
              Nuffield Department of Medicine, University of Oxford
}

\date{Received: date / Accepted: date}

\maketitle

\begin{abstract}
This paper examines methodology for performing Bayesian inference
sequentially on a sequence of posteriors on spaces of different dimensions.
For this we use sequential Monte
Carlo (SMC) samplers, introducing the innovation
of using deterministic transformations to move particles effectively
between target distributions with different dimensions. This approach,
combined with adaptive methods, yields an extremely flexible and general
algorithm for Bayesian model comparison that is suitable for use in
applications where the acceptance rate in reversible jump Markov chain
Monte Carlo (RJMCMC) is low. We use this approach on model comparison for mixture models, and for  inferring coalescent trees sequentially, as data arrives.
\keywords{Bayesian model comparison \and coalescent \and trans-dimensional Monte Carlo}
\end{abstract}

\section{Introduction}

\subsection{Sequential inference}

Much of the methodology for Bayesian computation is designed with
the aim of approximating a posterior $\pi$. The most prominent approach is to
use Markov chain Monte Carlo (MCMC), in which a Markov chain that
has $\pi$ as its limiting distribution is simulated. It is well known:
that this process may be computationally expensive; that it is not
straightforward to tune the method automatically; and that it can
be challenging to determine how long to run the chain for. Therefore,
designing and running an MCMC algorithm to sample from a particular
target $\pi$ may require much human input and computer time. This
creates particular problems if a user is in fact interested in a number
of target distributions $\left(\pi_{t}\right)_{t=1}^{T}$ defined possibly on different spaces: using MCMC
on each target requires additional computer time to run the separate
algorithms and each may require human input to design the algorithm,
determine the burn in, etc. This paper has as its subject the task
of using a Monte Carlo method to simulate from each of the targets
$\pi_{t}$ that avoids these disadvantages.

Particle filtering \citep{Gordon1993} and its generalisation, the
SMC sampler \citep{DelMoral2006c} is designed to tackle problems of this nature. Roughly speaking, the idea of these approaches is to begin
by using importance sampling (IS) to find a set of weighted \emph{particles}
that give an empirical approximation to $\pi_{0}$ then to, for $t=0,...,T-1$,
update the set of particles approximating $\pi_{t}$ such that they,
after changing their positions using a kernel $K_{t+1}$ and updating
their weights, approximate $\pi_{t+1}$. This approach is particularly
useful where neighbouring target distributions in the sequence are
similar to each other, and in this case has the following advantages
over running $T$ separate MCMC algorithms.
\begin{itemize}
\item The similarity of neighbouring targets can be exploited since particles
approximating $\pi_{t}$ may not need much adjustment to provide a
good approximation to $\pi_{t+1}$. We have the desirable property
that we find approximations to each of the targets in the sequence.
Further, we also may gain when compared to running a single MCMC algorithm
to target $\pi_{T}$, since it may be complicated to set up an MCMC
that simulates well from $\pi_{T}$ without using a sequence of simpler
distributions to guide particles into the appropriate regions of the
space.
\item When the targets $\left(\pi_{t}\right)_{t=1}^{T}$ are only known up to a constant of proportionality, SMC samplers also provide unbiased
estimates of the corresponding normalising constants. In a Bayesian context,
the normalising constant of $\pi_{t}$ is the \emph{marginal likelihood} or \emph{evidence}, a key quantity
in Bayesian model comparison. For much of the paper, and in abuse of notation, we use the same letters for denoting distributions and corresponding densities. In addition, we use tildes to denote unnormalised densities; e.g. let $\theta \sim \pi_t(\cdot)$ then its density is given by $\pi_t\left( \theta \right)=\tilde{\pi}_t\left( \theta \right)/Z_t$, where $Z_t$ denotes the normalising constant.
\end{itemize}

\subsection{Outline of paper\label{subsec:Targets-on-spaces}}

In this paper we consider the case where each $\pi_{t}$ is defined
on a space of different dimension, often of increasing dimension with
$t$. We provide a general framework for implementing an SMC algorithm in the aforementioned setting.
A particle filter is designed to be used in a special case of
this situation: the case where $\pi_{t}$ is the path distribution
in a state space model, $\pi_{t}\left(\theta_{1:t}\mid y_{1:t}\right)$.
A particle filter exploits the Markov property in order to update
a particle approximation of $\pi_{t}\left(\theta_{1:t}\mid y_{1:t}\right)$
to an approximation of $\pi_{t+1}\left(\theta_{1:t+1}\mid y_{1:t+1}\right)$.
In this paper we consider targets in which there is not such a straightforward
relationship between $\pi_{t}$ and $\pi_{t+1}$. In addition, the approach we present is useful in Bayesian model comparison that results from constructing
an SMC sampler where each $\pi_{t}$ corresponds to a different model and there
are $T$ models that can be ordered, usually in order of their complexity.
Deterministic transformations are used to move points between one
distribution and the next, potentially yielding efficient samplers by reducing
the distance between successive distributions. We also show how the
same framework can be used for sequential inference under the coalescent
model \citep{Kingman1982}.

The use of deterministic transformations to improve SMC has been considered
previously in a number of papers (e.g. \citet{Chorin2009,Vaikuntanathan2011,Reich2013,Heng2015,South2019}).
Several of these papers are focussed on how to construct useful transformations
in a generic way including, for example: methods that map high density
regions of the proposal to high density regions of the target \citep{Chorin2009};
and methods that approximate the solution of ordinary differential
equations that mimic the SMC dynamics \citep{Heng2015}. This paper
is different in that it focuses on the particular case of a sequence
of distribution on spaces of different dimensions, and uses transformations
and proposals that are designed for the applications we study.

Section \ref{sec:SMC-samplers-with} describes the methodology
introduced in the paper, considering both practical and theoretical
aspects, and provides comparison to existing methods. We provide an
example of the use of the methodology for Bayesian model comparison
in section \ref{sec:Bayesian-model-comparison}, on the Gaussian mixture
model. In section \ref{sec:Sequential-Bayesian-inference} we use
our methodology for online inference under the coalescent, using the
flexibility of our proposed approach to describe a method for moving
between coalescent trees. In section \ref{sec:Conclusions} we present a final discussion and outline possible extensions.

\section{SMC samplers with transformations\label{sec:SMC-samplers-with}}

\subsection{SMC samplers with increasing dimension}

The use of SMC samplers on a sequence of targets of increasing dimension
has been described previously (e.g. \citet{Naesseth2014,Everitt2017,Dinh2016}).
These papers introduce an additional proposal distribution for the
variables that are introduced at each step. In this section we straightforwardly see that this is a particular case of the SMC sampler in \citet{DelMoral2007}.

\subsubsection{SMC samplers with MCMC moves}

To introduce notation, we first consider the standard case in which
the dimension is fixed across all iterations of the SMC. For simplicity
we consider only SMC samplers with MCMC moves, and we consider an
SMC sampler that has $T$ iterations. Let $\pi_{t}$ be our target
distribution of interest at iteration $t$, this being the distribution
of the random vector $\theta_{t}$ on space $E$. Throughout the paper
the values taken by particles in the SMC sampler
have a $^{(p)}$ superscript to distinguish them from random vectors;
so for example $\theta_{t}^{(p)}$ is the value taken by the $p$th
particle. We define $\pi_{0}$ to be a distribution from which we
can simulate directly, simulate each particle $\theta_{0}^{(p)}\sim\pi_{0}$
and set its normalised weight $w_{0}^{(p)}=1/P$. Then for $0\leq t<T$
at the $\left(t+1\right)$th iteration of the SMC sampler, the following
steps are performed.
\begin{enumerate}
\item \textbf{Reweight}: Calculate the updated (unnormalised) weight $\tilde{w}_{t+1}^{(p)}$
of the $p$th particle
\begin{eqnarray}
\tilde{w}_{t+1}^{(p)} & = & w_{t}^{(p)}\frac{\tilde{\pi}_{t+1}\left(\theta_{t}^{(p)}\right)}{\tilde{\pi}_{t}\left(\theta_{t}^{(p)}\right)}.\label{eq:smc_weight}
\end{eqnarray}
\item \textbf{Resample: }Normalise the weights to obtain normalised weights
$w_{t+1}^{(p)}$ and calculate the \emph{effective sample size} (ESS)
\citep{Kong1994}. If the ESS falls below some threshold, e.g. $\alpha P$
where $0<\alpha<1$, then resample.
\item \textbf{Move: }For each particle use an MCMC move with target
$\pi_{t+1}$ to move $\theta_{t}^{(p)}$ to $\theta_{t+1}^{(p)}$.
\end{enumerate}
We remark that the move step above does not necessarily imply using a single MCMC iteration; if the chosen MCMC mixes slowly then performing many iterations and using adaptive strategies will result beneficial.
The previous algorithm yields an empirical approximation of $\pi_{t}$ and
an estimate of its normalising constant $Z_t$
\begin{equation}
\hat{\pi}_{t}^{P}=\sum_{p=1}^{P}w_{t}^{(p)}\delta_{\theta_{t}^{(p)}},\quad\widehat{Z}_t=\prod_{s=0}^{t}\sum_{p=1}^{P}w_{s}^{(p)}\frac{\tilde{\pi}_{s+1}\left(\theta_{s}^{(p)}\right)}{\tilde{\pi}_{s}\left(\theta_{s}^{(p)}\right)}\label{eq:smc_empirical_approx}
\end{equation}
where $\delta_{\theta}$ is a Dirac mass at $\theta$.

\subsubsection{Increasing dimension\label{subsec:Increasing-dimension}}

We now describe a case where the parameter $\theta$ increases in
dimension with the number of SMC iterations. Our approach is to set
up an SMC sampler on an extended space that has the same dimension
of the maximum dimension of $\theta$ that we will consider (similarly
to \citet{Carlin1995a}). At SMC iteration $t$, we use: $\theta_{t}$
to denote the random vector of interest; $u_{t}$ to denote a random
vector that contains the additional dimensions added to the parameter
space at iteration $t+1$, and $v_{t}$ to denote the remainder of
the dimensions that will be required at future iterations. Our SMC
sampler is constructed on a sequence of distributions $\varphi_{t}$
of the random vector $\vartheta_{t}=\left(\theta_{t},u_{t},v_{t}\right)$
in space $E=\left( \Theta_t, U_t, V_t \right)$, with
\begin{equation}
\varphi_{t}\left(\vartheta_{t}\right)=\pi_{t}\left(\theta_{t}\right)\psi_{t}\left(u_{t}\mid\theta_{t}\right)\phi_{t}\left(v_{t}\mid\theta_{t},u_{t}\right),\label{eq:smc_birth_target}
\end{equation}
where $\pi_{t}$ is the distribution of interest at iteration $t$,
and $\psi_{t}$ and $\phi_{t}$ are (normalised) distributions on
the additional variables so that $\pi_{t}$ and $\varphi_{t}$ have
the same normalising constant. The weight update in this SMC sampler
is
\begin{equation}
\tilde{w}_{t+1}^{(p)}=w_{t}^{(p)}\frac{\tilde{\pi}_{t+1}\left(\theta_{t}^{(p)},u_{t}^{(p)}\right)}{\tilde{\pi}_{t}\left(\theta_{t}^{(p)}\right)\psi_{t}\left(u_{t}^{(p)}\mid\theta_{t}^{(p)}\right)}.\label{eq:smc_birth}
\end{equation}
Here, as in particle filtering, by construction, the $\phi_{t}$ terms
in the numerator and denominator have cancelled so that none of the
dimensions added after iteration $t+1$ are involved; a characteristic
shared by the MCMC move with target $\varphi_{t+1}$, that need only
update $\theta_{t}$, $u_{t}$.

\subsection{Motivating example: Gaussian mixture models\label{subsec:Motivating-example:-Gaussian}}

\subsubsection{RJMCMC for Gaussian mixture models}

The following sections make use of transformations and other ideas
in order to improve the efficiency of the sampler. To motivate this,
we consider the case of Bayesian model comparison, in which the $\pi_{t}$
are different models ordered by their complexity. In section \ref{sec:Bayesian-model-comparison}
we present an application to Gaussian mixture models, and we use this
as our motivating example here.  Consider mixture models with $t$
components, to be estimated from data $y$, consisting of $N$ observed
data points. For simplicity we describe a ``without completion''
model, where we do not introduce a label $z$ that assigns data points
to components. Let the $s$th component have a mean $\mu_{s}$, precision
$\tau_{s}$ and weight $\nu_{s}$, with the weights summing to one
over the components. Let $p_{\mu}$ and $p_{\tau}$ be the respective
priors on these parameters, which are the same for every component,
and let $p_{\nu}$ be the joint prior over all of the weights. The
likelihood under $t$ components is
\begin{equation}
f_{t}\left(y\mid\theta_{t}=\left(\mu_{s},\tau_{s},\nu_{s}\right)_{s=1}^{t}\right)=\prod_{i=1}^{N}\sum_{s=1}^{t}\nu_{s}\mathcal{N}\left(y_{i}\mid\mu_{s},\tau_{s}^{-1}\right).\label{eq:gmm_llhd}
\end{equation}

An established approach for estimating mixture models is that of RJMCMC.
Here, $t$ is chosen to be a random variable and assigned a prior
$p_{t}$, which here we choose to be uniform over the values $1$
to $T$. Let
\begin{align}
&\pi_{t}\left(\theta_{t}\right)=\pi\left(\theta_{t}\mid t,y\right) \nonumber \\
&\quad\propto p_{\nu}\left(\nu_{1:t}\right)\left(\prod_{s=1}^{t}p_{\mu}\left(\mu_{s}\right)p_{\tau}\left(\tau_{s}\right)\right)f_{t}\left(y\mid\theta_{t}=\left(\mu_{s},\tau_{s},\nu_{s}\right)_{s=1}^{t}\right)\label{eq:gmm_pi}
\end{align}
be the joint posterior distribution over the parameters $\theta_{t}$
conditional on $t$. RJMCMC simulates from the joint space of $\left(t,\theta_{t}\right)$
in which a mixture of moves is used, some fixed-dimensional ($t$
fixed) and some trans-dimensional (to mix over $t$). The simplest
type of trans-dimensional move in this case is that of a birth
move for moving from $t$ to $t+1$ components, or a death move for
moving from $t+1$ to $t$ \citep{Richardson1997}. We consider
a birth move, a uniform prior probability over
$t$ and equal probability of proposing birth or death.  For the purposes of exposition we assume that the
weights of the components are chosen to be fixed in each model (this
assumption will be relaxed later in section \ref{sec:Bayesian-model-comparison}). Let $u_{t}=\left(\mu_{t+1},\tau_{t+1}\right)$,
be the mean and precision of the new component and let $\psi_{t}\left(u_{t}\mid\theta_{t}\right)=p_{\mu}\left(\mu_{t+1}\right)p_{\tau}\left(\tau_{t+1}\right)$.
A birth move simulates $u_{t}\sim\psi_{t}$ and has acceptance probability
\begin{equation}
\alpha=\min\left\{ 1,\frac{\pi_{t+1}\left(\theta_{t+1}\right)}{\pi_{t}\left(\theta_{t}\right)\psi_{t}\left(u_{t} \mid \theta_t \right)}\right\},\label{eq:rj_birth_acceptance}
\end{equation}
where $\theta_{t+1}=\left( \theta _t, u_t \right)$.

\subsubsection{Comparing RJMCMC and SMC samplers\label{subsec:Linking-RJMCMC-and}}

Consider the use of an SMC sampler for inference where the sequence
of target distributions is $\left(\pi_{t}\right)_{t=1}^{T}$, i.e.
the $t$th distribution is the mixture of Gaussians with $t$ components.
By choosing $u_t$ and $\psi_t$ as above, together with 
\begin{align*}
v_{t}=\left(\mu_{(t+2):T},\tau_{(t+2):T}\right)
\end{align*}
and 
\begin{align*}
\phi_{t}\left(v_{t}\mid\theta_{t},u_{t}\right)=\prod_{s=t+2}^{T}p_{\mu}\left(\mu_{s}\right)p_{\tau}\left(\tau_{s}\right),
\end{align*}
we may use the SMC sampler described in section \ref{subsec:Increasing-dimension}.
Note that the ratio in the acceptance probability in equation (\ref{eq:rj_birth_acceptance})
is the same as the incremental SMC weight in equation (\ref{eq:smc_birth}).
The reason for this is that both algorithms make use of an IS estimator
of the Bayes factor $Z_{t+1}/Z_{t}$: using a proposed point $\theta_{t}\sim\pi_{t}$,
$u_{t}\sim\psi_{t}$ and $\theta_{t+1}=\left(\theta_{t},u_{t}\right)$,
this estimator is given by
\begin{equation}
\widehat{\frac{Z_{t+1}}{Z_{t}}}=\frac{\pi_{t+1}\left(\theta_{t+1}\right)}{\pi_{t}\left(\theta_{t}\right)\psi_{t}\left(u_{t}\mid\theta_{t}\right)}.\label{eq:rjmcmc_is}
\end{equation}
We may see RJMCMC as using an IS estimator of the ratio of the posterior
model probabilities within its acceptance ratio; this view on RJMCMC
\citep{Karagiannis2013} links it to pseudo-marginal approaches \citep{Andrieu2009}
in which IS estimators of target distributions are employed. As in
pseudo-marginal MCMC, the efficiency of the chain depends on the variance
of the estimator that is used. We observe that the IS estimator in
equation (\ref{eq:rjmcmc_is}) is likely to have high variance: this
is one way of explaining the poor acceptance rate of dimension changing
moves in RJMCMC. In particular, we note that this estimator suffers
a curse of dimensionality in the dimension of $\theta_{t+1}$, meaning that RJMCMC is
in practice seldom effective when the parameter space is of high dimension.
This view suggests a number of potential improvements to RJMCMC with
a birth move, each of which has been previously investigated.
\begin{itemize}
\item IS performs better if the proposal distribution is close to the target,
whilst ensuring that the proposal has heavier tails than the target.
The original RJMCMC algorithm allows the possibility to construct
such proposals by allowing for the use of transformations to move
from the parameters of one model to the parameters of another. \citet{Richardson1997}
provide a famous example of this in the Gaussian mixture case in the
form of split-merge moves. Focusing on the split move, the idea is
to propose splitting an existing component, using a moment matching
technique to ensure that the new components have appropriate means,
variances and weights.
\item Annealed importance sampling (AIS) \citep{Neal2001} yields a lower variance than IS. The idea is to use intermediate
distributions to form a path between the IS proposal and target, using
MCMC moves to move points along this path. This approach was shown
to be beneficial in some cases by \citet{Karagiannis2013}.
\item The estimator in equation (\ref{eq:rjmcmc_is}) uses only a single importance
point. It would be improved by using multiple points. However, using
such an estimator directly within RJMCMC leads to a ``noisy'' algorithm
that does not have the correct target distribution for the same reasons
as those given for the noisy exchange algorithm in \citet{Alquier2016}.
We note that recent work \citep{Andrieu2018} suggests a correction
to provide an exact approach based on the same principle.
\end{itemize}
The approach we take in this paper is to investigate variations on
these ideas within the SMC sampler context, rather than RJMCMC. We
begin by examining the use of transformations in section \ref{subsec:Using-transformations-in},
then describe the use of intermediate distributions and other refinements
in section \ref{subsec:Design-of-SMC}. The final idea is automatically
used in the SMC context, due to the use of $P$ particles.

\subsection{Using transformations in SMC samplers\label{subsec:Using-transformations-in}}

We now show (in a generalisation of section
\ref{subsec:Increasing-dimension}) how to use transformations within SMC,
whilst simultaneously changing the dimension of the target at each
iteration; an approach we will refer to as \emph{transformation SMC
}(TSMC). We again use the approach of performing
SMC on a sequence of targets $\varphi_{t}$, with each of the these
targets being on a space of fixed dimension, constructed such that
they have the desired target $\pi_{t}$ as a marginal. In this section
the dimension of the space on which $\pi_{t}$ is defined again varies
with $t$, but is not necessarily increasing with $t$. Let $\theta_{t}$
be the random vector of interest at SMC iteration $t$: we wish to
approximate the distributions $\pi_{t}$ of $\theta_{t}$ in the space
$\Theta_{t}$. Let $\left( \tilde{\varphi}_{t} \right)_{t=1}^T$ be a sequence of unnormalised targets, whose normalised versions are $\left( \varphi_{t} \right)_{t=1}^T$ and being the distribution
of the random vector $\vartheta_{t}=\left(\theta_{t},u_{t}\right)$
in the space $E_{t}=\left(\Theta_{t},U_{t}\right)$ where
\[
\tilde{\varphi}_{t}\left(\theta_{t},u_{t}\right)=\tilde{\pi}_{t}\left(\theta_{t}\right)\psi_{t}\left(u_{t}\mid\theta_{t}\right),
\]
implying $\varphi_{t}$ and $\pi_{t}$ have the same normalising constant $Z_t$.
The dimension of $\Theta_{t}$ can change with $t$, but the dimension
of $E_{t}$ must be constant in $t$. We introduce a transformation
$G_{t\rightarrow t+1}:\Theta_{t}\times U_{t}\rightarrow\Theta_{t+1}\times U_{t+1}$
and define
\[
\vartheta_{t\rightarrow t+1}=\left(\theta_{t\rightarrow t+1}\left(\vartheta_{t}\right),u_{t\rightarrow t+1}\left(\vartheta_{t}\right)\right):=G_{t\rightarrow t+1}\left(\vartheta_{t}\right).
\]
In many cases we will choose $G_{t\rightarrow t+1}$ to be bijective.
In this case we denote its inverse by $G_{t+1\rightarrow t}=G_{t\rightarrow t+1}^{-1}$,
with
\begin{align*}
\vartheta_{t+1\rightarrow t}&=\left(\theta_{t+1\rightarrow t}\left(\vartheta_{t+1}\right),u_{t+1\rightarrow t}\left(\vartheta_{t+1}\right)\right)\\
&:=G_{t+1\rightarrow t}\left(\vartheta_{t+1}\right).
\end{align*}
Let the distribution of the transformed random variable $\vartheta_{t\rightarrow t+1}$
be $\varphi_{t\rightarrow t+1}$, i.e. $\varphi_{t\rightarrow t+1}=\mathcal{L}\left(\vartheta_{t\rightarrow t+1}\right)=\mathcal{L}\left(G_{t\rightarrow t+1}\left(\vartheta_{t}\right)\right)$
where $\mathcal{L}\left(X\right)$ denotes the law of a random variable
$X$, and let the distribution of $\vartheta_{t+1\rightarrow t}$
be $\varphi_{t+1\rightarrow t}$. These distributions may be derived
using standard results about the distributions of transforms of random
variables: e.g. where the $E_{t}$ are continuous spaces and
where $G_{t\rightarrow t+1}$ is a diffeomorphism, having Jacobian
determinant $J_{t\rightarrow t+1}$ , with inverse $G_{t+1\rightarrow t}$
having Jacobian determinant $J_{t+1\rightarrow t}$. In this case we
have
\begin{align*}
&\tilde{\varphi}_{t\rightarrow t+1}\left(\vartheta_{t\rightarrow t+1}\right)  =  \tilde{\pi}_{t}\left(\theta_{t+1\rightarrow t}\left(\vartheta_{t\rightarrow t+1}\right)\right) \\
&\quad \times \psi_{t}\left(u_{t+1\rightarrow t}\left(\vartheta_{t\rightarrow t+1}\right)\mid\theta_{t+1\rightarrow t}\left(\vartheta_{t\rightarrow t+1}\right)\right)\left|J_{t+1\rightarrow t}\right|,
\end{align*}
\begin{align*}
& \tilde{\varphi}_{t+1\rightarrow t}\left(\vartheta_{t+1\rightarrow t}\right) =  \tilde{\pi}_{t+1}\left(\theta_{t\rightarrow t+1}\left(\vartheta_{t+1\rightarrow t}\right)\right) \\
&\quad \times \psi_{t+1}\left(u_{t\rightarrow t+1}\left(\vartheta_{t+1\rightarrow t}\right)\mid\theta_{t\rightarrow t+1}\left(\vartheta_{t+1\rightarrow t}\right)\right)\left|J_{t\rightarrow t+1}\right|.
\end{align*}
We may then use an SMC sampler on the sequence of targets $\varphi_{t}$,
with the following steps at its $(t+1)$th iteration.
\begin{enumerate}
\item \textbf{Transform: }For the $p$th particle, apply\textbf{ }$\vartheta_{t\rightarrow t+1}^{(p)}=G_{t\rightarrow t+1}\left(\vartheta_{t}^{(p)}\right)$.
\item \textbf{Reweight and resample}: Calculate the updated (unnormalised)
weight $\tilde{w}_{t+1}^{(p)}$
\begin{eqnarray}
\tilde{w}_{t+1}^{(p)} & = & w_{t}^{(p)}\frac{\tilde{\varphi}_{t+1}\left(\vartheta_{t\rightarrow t+1}^{(p)}\right)}{\tilde{\varphi}_{t\rightarrow t+1}\left(\vartheta_{t\rightarrow t+1}^{(p)}\right)}.\label{eq:smc_trans_weight2}
\end{eqnarray}
Where $G_{t\rightarrow t+1}$ is a diffeomorphism we have
\begin{equation}
\tilde{w}_{t+1}^{(p)}=w_{t}^{(p)}\frac{\tilde{\pi}_{t+1}\left(\theta_{t\rightarrow t+1}^{(p)}\right)\psi_{t+1}\left(u_{t\rightarrow t+1}^{(p)}\mid\theta_{t\rightarrow t+1}^{(p)}\right)}{\tilde{\pi}_{t}\left(\theta_{t}^{(p)}\right)\psi_{t}\left(u_{t}^{(p)}\mid\theta_{t}^{(p)}\right)\left|J_{t+1\rightarrow t}\right|}.\label{eq:tsmc_weight_jacobian}
\end{equation}
It is possible, depending on the transformation used, that this weight
update involves none of the dimensions above $\max\left\{ \dim\left(\theta_{t}\right),\dim\left(\theta_{t+1}\right)\right\} $ as happened in \eqref{eq:smc_birth}.
Then resample if the ESS falls below some threshold, as described
previously.
\item \textbf{Move. }For each $p$, let $\vartheta_{t+1}^{(p)}$ be the result of an MCMC move with target
$\varphi_{t+1}$, starting from $\vartheta_{t\rightarrow t+1}^{(p)}$. We need not simulate $u$ variables that are not used at the next iteration.
\end{enumerate}
To illustrate the additional flexibility this framework allows, over
and above the sampler described in section \ref{subsec:Increasing-dimension},
we consider the Gaussian mixture example in section \ref{subsec:Motivating-example:-Gaussian}.
The sampler from \ref{subsec:Increasing-dimension} provides an alternative
to RJMCMC in which a set of particles is used to sample from each
model in turn, using the particles from model $t$, together with
new dimensions simulated using a birth move, to explore model $t+1$.
The sampler in this section allows us to use a similar idea using
more sophisticated proposals, such as split moves.
The efficiency of the sampler depends on the choice of $\psi_{t}$
and $G_{t\rightarrow t+1}$. As previously, a good choice for these
quantities should result in a small distance between $\varphi_{t\rightarrow t+1}$
and $\varphi_{t+1}$, whilst ensuring that $\varphi_{t\rightarrow t+1}$
has heavier tails than $\varphi_{t+1}$.
As in the design of RJMCMC algorithms, usually these choices will
be made using application-specific insight.

\subsection{Design of SMC samplers\label{subsec:Design-of-SMC}}

\subsubsection{Using intermediate distributions\label{subsec:Using-intermediate-distributions}}

The Monte Carlo variance of an SMC sampler depends on the distance
between successive target distributions,
thus a well designed sampler will use a sequence of distributions
in which the distance between successive distributions is small. We
ensure this by introducing intermediate distributions in between successive
targets \citep{Neal2001}: in between targets $\varphi_{t}$ and $\varphi_{t+1}$
we use $K-1$ intermediate distributions, the $k$th being $\varphi_{t,k}$,
so that $\varphi_{t,0}=\varphi_{t}$ and $\varphi_{t,K}=\varphi_{t+1}$
and therefore $\varphi_{t,K}=\varphi_{t+1,0}$. We use \emph{geometric
annealing}, i.e.\emph{
\begin{align}
& \tilde{\varphi}_{t\rightarrow t+1,k}\left(\vartheta_{t\rightarrow t+1,k}\right)  \nonumber \\
&\quad =  \left[\tilde{\varphi}_{t+1}\left(\vartheta_{t\rightarrow t+1,k}\right)\right]^{\gamma_{k}}\left[\tilde{\varphi}_{t\rightarrow t+1}\left(\vartheta_{t\rightarrow t+1,k}\right)\right]^{1-\gamma_{k}},\label{eq:geom2}
\end{align}
}where $0=\gamma_{0}<...<\gamma_{K}=1$. This idea results in only
small alterations to the TSMC presented above. We now use a sequence
of targets $\varphi_{t,k}$, incrementing the $t$ index when $k=K$, then setting $k=0$ and finally using
a transform move $\vartheta_{t\rightarrow t+1,0}^{(p)}=G_{t\rightarrow t+1}\left(\vartheta_{t,K}^{(p)}\right)$ for
each $p\in \left\{ 1,\dots,P \right\}$. The weight update becomes
\begin{eqnarray}
\tilde{w}_{t,k+1}^{(p)} & = & w_{t,k}^{(p)}\frac{\tilde{\varphi}_{t\rightarrow t+1,k+1}\left(\vartheta_{t\rightarrow t+1,k}^{(p)}\right)}{\tilde{\varphi}_{t\rightarrow t+1,k}\left(\vartheta_{t\rightarrow t+1,k}^{(p)}\right)},\label{eq:smc_trans_weight2_intermediate}
\end{eqnarray}
and the MCMC moves now have target $\varphi_{t\rightarrow t+1,k+1}$,
starting from $\vartheta_{t\rightarrow t+1,k}^{(p)}$ and storing
the result in $\vartheta_{t\rightarrow t+1,k+1}^{(p)}$. The use of
intermediate distributions makes this version of TSMC more robust
than the previous one; the MCMC moves used at the intermediate distributions
provide a means for the the algorithm to recover if the initial transformation
is not enough to ensure that $\varphi_{t\rightarrow t+1}$ is similar
to $\varphi_{t+1}$.

\subsubsection{Adaptive SMC\label{subsec:Adapting-the-sequence}}

Section \ref{subsec:Using-intermediate-distributions} describes the
use of intermediate distributions with the aim of ensuring that the
distance between neighbouring targets is not too great, but this aim
cannot be achieved without also considering where to place these intermediate
distributions. In
this paper we follow the adaptive strategy used in \citet{Jasra2011,DelMoral2012g}
and refined in \citet{Zhou2015} in the case where resampling is not
performed at every iteration. At iteration $t$, $\left(k+1\right)$
this approach uses the conditional ESS (CESS)
\begin{equation}
\text{CESS}_{t,k+1}=\frac{P\left(\sum_{p=1}^{P}w_{t,k}^{(p)}\omega^{(p)}\right)^{2}}{\sum_{p=1}^{P}w_{t,k}^{(p)}\left(\omega^{(p)}\right)^{2}},\label{eq:cess}
\end{equation}
to monitor the discrepancy between neighbouring distributions, where
$\omega^{(p)}$ is the incremental weight given by the ratio multiplying $w_{t,k}^{(p)}$ in \eqref{eq:smc_trans_weight2_intermediate}.
Before the reweighting step is performed, the next intermediate distribution
is chosen to be the distribution under which the CESS is found to
be $\beta P$, for some $0<\beta<1$. In the case of the geometric
annealing scheme, this corresponds
to a particular choice for $\gamma_{k}$ for computing \eqref{eq:geom2}. As commented previously, we may also adapt the MCMC kernels used for the move step, based on the current particle set. For the two examples presented later, we have considered adaptive and non-adaptive strategies in the MCMC kernels. We refer the interested reader to the supplementary material for the specific details. Algorithm \ref{alg:TSMC} presents a generic version of TSMC using adaptive resampling and number of intermediate distributions.

\subsubsection{Auxiliary variables in proposals}

For the Gaussian mixture example, for two or more components, when
using a split move we must choose the component that is to be
split. We may think of the choice of splitting different components
as offering multiple ``routes'' through a space of distributions,
with the same start and end points. Another alternative route would
be given by using a birth move rather than a split move. In this section
we generalise TSMC to allow multiple routes. We restrict our attention
to the case where the choice of multiple routes is possible at the
beginning of a transition from $\varphi_{t}$ to $\varphi_{t+1}$,
when $k=0$ (more general schemes are possible). A route corresponds to a particular choice for the
transformation $G_{t\rightarrow t+1}$, thus we consider a set of
$M_{t}$ possible transformations indexed by the discrete random variable $l_{t}$, using the notation
$G_{t\rightarrow t+1}^{\left(l_{t}\right)}$ (also using this superscript
on distributions that depend on this choice of $G$). We now augment
the target distribution with variables $l_{0},...,l_{T-1}$ and, for
each $t$ alter the distribution $\psi_{t}$ such that it becomes
a joint distribution on $u_{t}$ and $l_{t}$. Our sampler will draw
the $l$ variables at the point at which they are introduced, so that
different particles use different routes, but will not perform any
MCMC moves on the variable after it is introduced. This leads to the
sampler being degenerate in most of the $l$ variables, but this doesn't
affect the desired target distribution.

A revised form of TSMC is then, when $k=0$, to first simulate routes
$l_{t}^{(p)}\sim\rho_{t}$ for each particle, then to use a different
transform $\vartheta_{t\rightarrow t+1,0}^{(p)}=G_{t\rightarrow t+1}^{\left(l_{t}^{(p)}\right)}\left(\vartheta_{t,K}^{(p)}\right)$
dependent on the route variable. The weight update is then given by
\begin{equation}
\tilde{w}_{t+1}^{(p)}=w_{t}^{(p)}\frac{\tilde{\pi}_{t+1}\left(\theta_{t\rightarrow t+1}^{(p)}\right)\psi_{t+1}\left(u_{t\rightarrow t+1}^{(p)},l_{t}^{(p)}\mid\theta_{t\rightarrow t+1}^{(p)}\right)}{\tilde{\pi}_{t}\left(\theta_{t}^{(p)}\right)\psi_{t}\left(u_{t}^{(p)},l_{t}^{(p)}\mid\theta_{t}^{(p)}\right)\left|J_{t+1\rightarrow t}^{\left(l_{t}^{(p)}\right)}\right|},\label{eq:tsmc_weight_route}
\end{equation}
where for simplicity we have omitted the dependence of $u_{t}^{(p)}$,
$u_{t\rightarrow t+1}^{(p)}$ and $\theta_{t\rightarrow t+1}^{(p)}$
on $l_{t}^{(p)}$. This weight update is very similar to one found
in \citet{DelMoral2006c}, for the case where a discrete auxiliary
variable is used to index a choice of MCMC kernels used in the move
step. Analogous to \citet{DelMoral2006c},
the variance of (\ref{eq:tsmc_weight_route}) is always greater than
or equal to that of (\ref{eq:tsmc_weight_jacobian}); we present an example in section
\ref{sec:Bayesian-model-comparison} where this additional variance
can result in large errors in marginal likelihood estimates). Alternatively one can employ the Rao-Blackwellisation procedure found in population
Monte Carlo \citep{Douc2007a}, and marginalise the proposal over
the auxiliary variable $l_{t}$. This results in a weight update of
\begin{equation}
\tilde{w}_{t+1}^{(p)}=w_{t}^{(p)}\frac{\pi_{t+1}\left(\theta_{t\rightarrow t+1}^{(p)}\right)\psi_{t+1}\left(u_{t\rightarrow t+1}^{(p)}\mid\theta_{t\rightarrow t+1}^{(p)}\right)}{\sum_{m=1}^{M_{t}}\pi_{t}\left(\theta_{t}^{(p)}\right)\psi_{t}\left(u_{t}^{(p)},m\mid\theta_{t}^{(p)}\right)\left|J_{t+1\rightarrow t}^{\left(m\right)}\right|}.\label{eq:tsmc_weight_route_rb}
\end{equation}
As mentioned in \citet{DelMoral2006c}, using \eqref{eq:tsmc_weight_route_rb} comes with extra computational cost, which could be prohibitively large if $M_t$ is large.

\begin{algorithm}[h]
\DontPrintSemicolon
  
  \KwInput{Particle approximation $\{ \vartheta_{t}^{(p)}, w_t^{(p)}\}_{p=1}^{P}\approx \varphi_t$; ESS threshold $\alpha \in (0,1)$; CESS threshold $\beta \in (0,1)$. }
  \KwOutput{Particle approximation $\{ \vartheta_{t+1}^{(p)}, w_{t+1}^{(p)}\}_{p=1}^{P}\approx \varphi_{t+1}$; Estimator $\widehat{Z_{t+1}/Z_t}$. }
  
  Initialise $k=0$, $\gamma_k=0$, $Z:=\widehat{Z_{t+1}/Z_t}=1$.\;
  \ForEach{$p \in \{ 1,\dots,P \}$}{
  Transform particle $\vartheta_{t\rightarrow t+1,k}^{(p)}=G_{t\rightarrow t+1}( \vartheta_{t}^{(p)})$.\;
  Set $w_{t,k}^{(p)}=w_t^{(p)}$.\;
  }
  \While{$\gamma_k<1$}{
  
  Find $\gamma_{k+1} \in (0,1]$ such that $CESS_{t,k+1}=\beta P$.\;
  
  \ForEach{$p \in \{ 1,\dots,P \}$}{
  
  Compute the weight $\tilde{w}_{t,k+1}^{(p)}$ using \eqref{eq:smc_trans_weight2_intermediate}.\;
  }
  Update $Z=Z\sum_{p=1}^{P} \tilde{w}_{t,k+1}^{(p)}$.\;
  Renormalise the above weights to obtain $\{ w_{t,k+1}^{(p)} \}_{p=1}^P$. \;
  
  \If{$ESS_{t,k+1} < \alpha P $}{
  		
    		Resample particles an set $w_{t,k+1}^{(p)}=1/P$ for all $p \in \{ 1,\dots,P \}$.\;
	}
	
	\ForEach{$p \in \{ 1,\dots,P \}$}{
  Set $\vartheta_{t\rightarrow t+1,k+1}^{(p)}\sim \mathcal{K}_{t,k+1}(\vartheta_{t\rightarrow t+1,k}^{(p)}, \cdot )$, where $\mathcal{K}_{t,k+1}$ is a $\varphi_{t\rightarrow t+1,k+1}$-invariant MCMC kernel.
  }
  Set $k=k+1$.\;
  
  }

\caption{TSMC algorithm with adaptive resampling and intermediate distributions}
\label{alg:TSMC}
\end{algorithm}

\subsection{Discussion}

One of the most obvious applications of TSMC is Bayesian model comparison.
SMC samplers are a generalisation of several other techniques, such
as IS, AIS and the ``stepping stone'' algorithm from \citet{Xie2011}
(which is essentially performing AIS with a reversed sequence of distributions, with more than one MCMC move
is used per target distribution), thus we expect a well-designed SMC
to outperform these techniques in most cases. \citet{Zhou2015} reviews
existing techniques that use SMC for model comparison, and concludes
that ``the SMC2 algorithm (moving from prior to posterior) with adaptive
strategies is the most promising among the SMC strategies''. In section
\ref{sec:Bayesian-model-comparison} we provide a detailed comparison
of TSMC with SMC2, and find that TSMC can have significant advantages.

Section \ref{subsec:Linking-RJMCMC-and} compared TSMC with RJMCMC,
noting that RJMCMC explores the model space by using a high variance
estimator of a Bayes factor at each MCMC iteration, whereas TSMC is
designed to construct a single lower variance estimator of each Bayes
factor. The high variance estimators within RJMCMC are the cause of
its most well known drawback: that the acceptance rate of trans-dimensional
moves can be very small. The design of TSMC, in which each model is
visited in turn, completely avoids this issue. One might envisage
that despite avoiding poor mixing, TSMC might instead yield high variance
Bayes factor estimators for challenging problems. However, TSMC has
the advantage that that adaptive methods may be used in order to reduce
the possibility that the estimators have high variance by, for example,
automatically using more intermediate distributions. The possibility
to adaptively choose intermediate distributions also provides an advantage
over the approach of \citet{Karagiannis2013}, where a sequence of
intermediate distributions for estimating each Bayes factor must be
specified in advance.

Since, by construction, TSMC is a particular instance of SMC as described in \citet{DelMoral2006c}, all of the theoretical properties of a standard SMC algorithm apply. Of particular interest are the properties of the method as the dimension of the parameter spaces grows. TSMC is constructed on a sequence of extended spaces $E_t$, each of which has dimension $d_{T}$, thus in the worst case, the results for an SMC sampler on a space of dimension $d_{T}$ apply. In this respect, the authors in \citet{Beskos2014b} have analysed
the stability of SMC samplers as the dimension of the state-space
increases when the number of particles $P$ is fixed. Their work provides
justification, to some extent, for the use of intermediate distributions
$\left( \varphi_{t,k}\right)_{k=1}^{K}$. Under fairly strong assumptions,
it has been shown that when the number of intermediate distributions
$K=\mathcal{O}\left(d_{T}\right)$, and as $d_{T}\rightarrow\infty$,
the effective sample size $\mbox{ESS}_{t+1}^{P}$ is stable in the
sense that it converges to a non-trivial random variable taking values
in $\left(1,P\right)$. The total computational cost for bridging
$\varphi_{t}$ and $\varphi_{t+1}$, assuming a product
form of $d_{T}$ components, is $\mathcal{O}\left(Pd_{T}^{2}\right)$. However, in practice, due to the cancellation of ``fill in'' variables, and using sensible transformations between consecutive distributions, one could expect a much lower effective dimension of the problem; an example of this situation is presented in the next section. Some theoretical properties of the method are explored further in the Supplementary Information.

\section{Bayesian model comparison for mixtures of Gaussians\label{sec:Bayesian-model-comparison}}

In this section we examine the use of TSMC on the mixture of Gaussians
application in section \ref{subsec:Motivating-example:-Gaussian}:
i.e. we wish to perform Bayesian inference of the number of components
$t$, and their parameters $\theta_{t}$, from data $y$. For simplicity,
we study the ``without completion'' model, where component labels
for each measurement are not included in the model. In the next sections
we outline the design of the algorithms used, then in section \ref{subsec:Results}
we describe the results of using these approaches on previously studied
data, highlighting features of the approach. Further results are given in the Supplementary Information.

\subsection{Description of algorithms\label{subsec:Description-of-algorithms}}

Let $t$ be the unknown number of mixture components, and $\left(\mu_{1:t},\tau_{\text{1:t}},\nu_{1:t}\right)$
(means, precisions and weights respectively) be the parameters of
the $t$ components. Our likelihood is the same as in equation (\ref{eq:gmm_llhd});
we use priors $\tau\sim\text{Gamma}\left(2,2S^{2}/100\right),$ $\nu_{1:t}\sim\text{Dir}\left(1,...,1\right)$
for the precisions and weights respectively, and for the means we
choose an unconstrained prior of $\mu\sim\mathcal{N}\left(m,S^{2}\right)$,
where $m$ is the mean and $S$ is the range of the observed data.
We impose an ordering constraint on the means, as described in \citet{Jasra2005e},
which simplifies the problem by eliminating many posterior modes with the added benefit of improving the interpretability of our results.
For simplicity we have also not included the commonly used ``random
beta'' hierarchical prior structure on $\tau$ \citep{Richardson1997}, which
from a statistical perspective is suboptimal but which simplifies our presentation of the behaviour of TSMC.

We use different variants of TSMC (as described in section \ref{subsec:Using-transformations-in}),
using a sequence of distributions $\left(\varphi_{t}\right)_{t=1}^{T}$
where $\varphi_{t}\left(\vartheta_{t}\right)=\pi_{t}\left(\theta_{t}\right)\psi_{t}\left(u_{t}\right)$.
$\pi_{t}$ is here the posterior on $t$ components given by equation
\ref{eq:gmm_pi}, and $\psi_{t}$ is different depending on the transformation
that is chosen. We use intermediate distributions (as described in
section \ref{subsec:Using-intermediate-distributions}), using geometric
annealing, in all of our algorithms, making use of the adaptive method
from section \ref{subsec:Adapting-the-sequence} to choose how to
place these distributions. The results in this section focus particularly
on illustrating the advantages afforded by making an intelligent choice
of the transformation in TSMC. Full details of the transformations, weight updates and MCMC moves are given in the Supplementary Information. In summary, we use the birth and split moves referred to in section \ref{subsec:Motivating-example:-Gaussian}, together with a move that orders the components. For both moves we present results using the weight updates in equations (\ref{eq:tsmc_weight_route}) (referred to henceforth as the conditional approach) and (\ref{eq:tsmc_weight_route_rb}) (referred to as the marginal approach).

\subsection{Results\label{subsec:Results}}

\begin{figure*}
\begin{tabular}{cc}

\begin{subfigure}[t]{0.5\textwidth}
        \centering
        \includegraphics[width=0.9\textwidth]{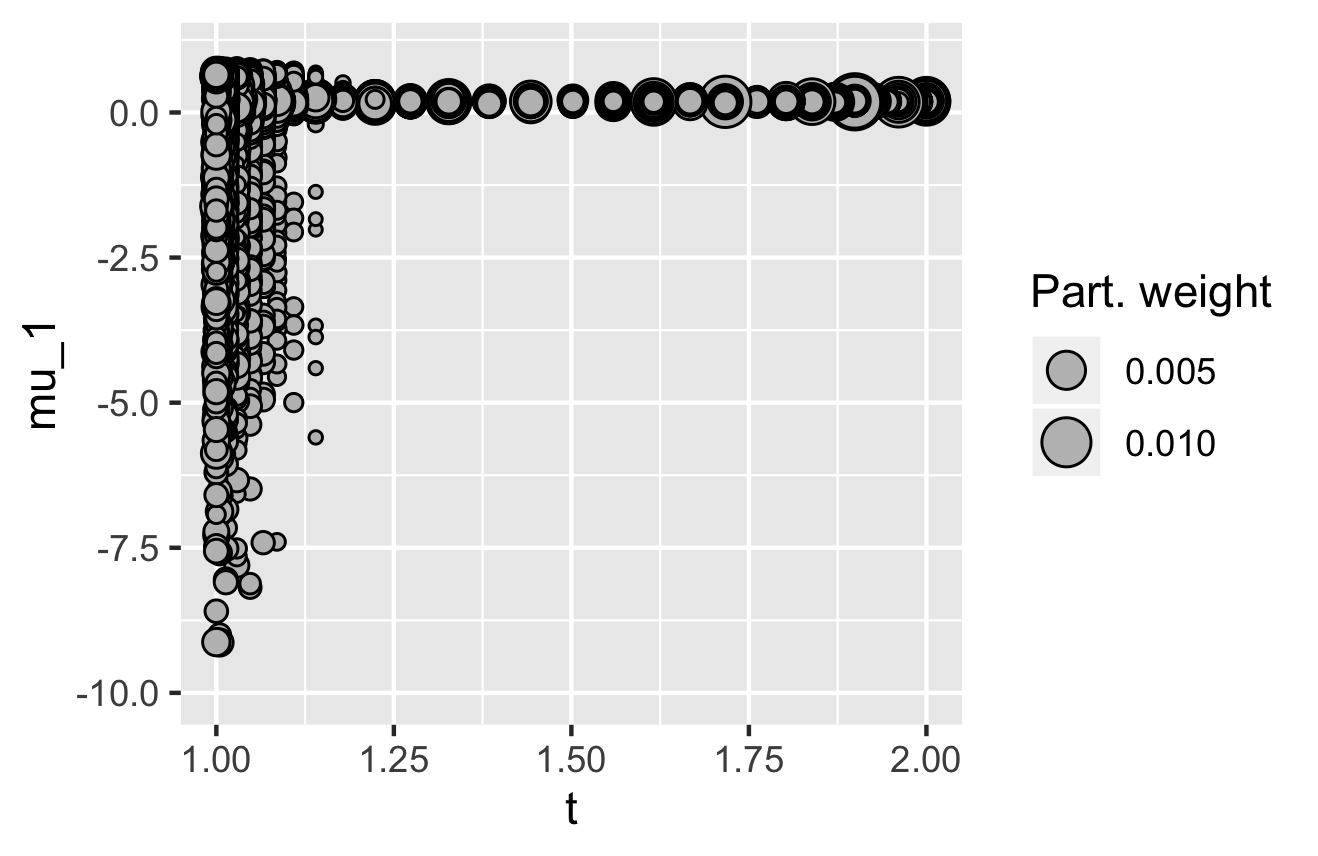}
        \caption{$\mu_{1}$ (birth).}
\end{subfigure}

\begin{subfigure}[t]{0.5\textwidth}
        \centering
        \includegraphics[width=0.9\textwidth]{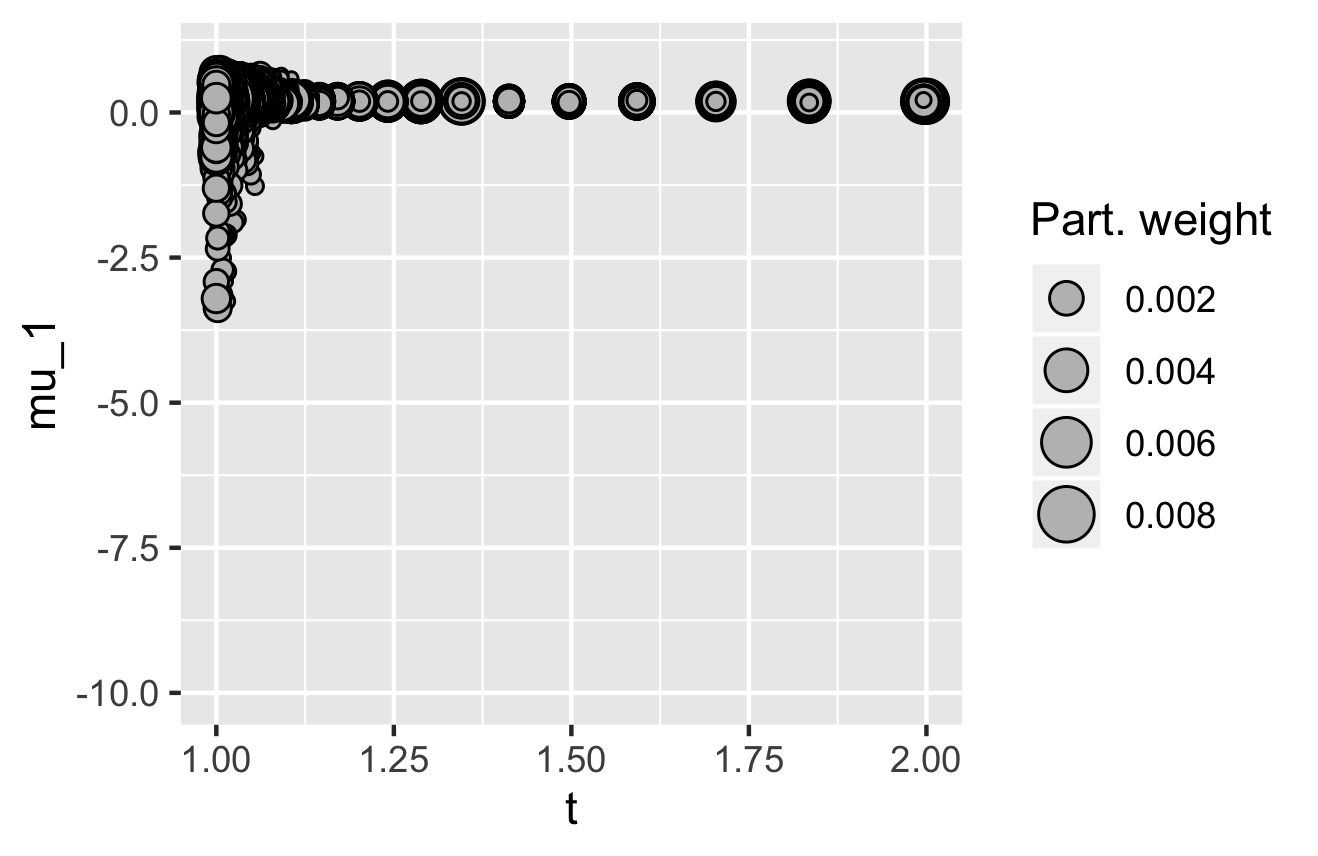}
        \caption{$\mu_{1}$ (split).}
\end{subfigure}

\\

\begin{subfigure}[t]{0.5\textwidth}
        \centering
        \includegraphics[width=0.90\textwidth]{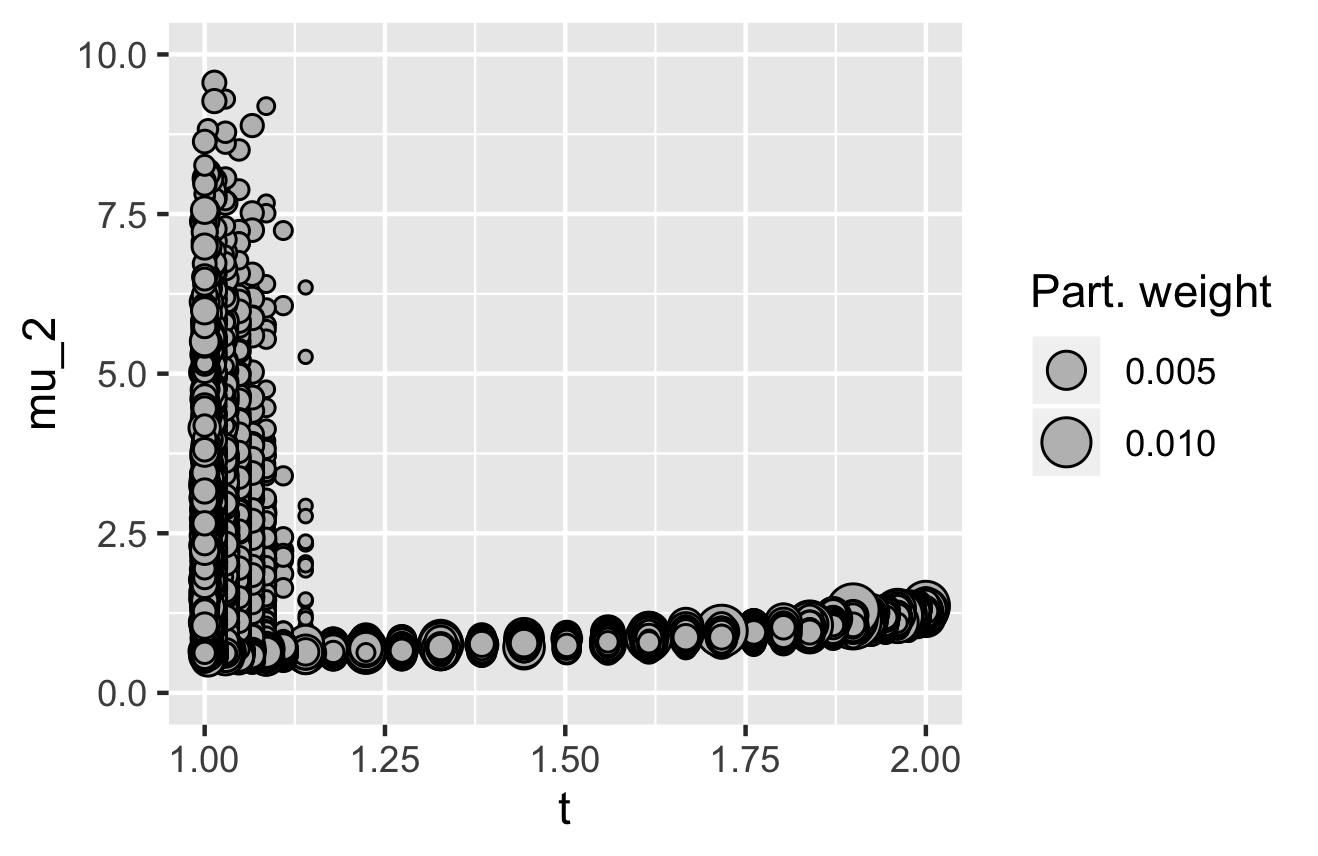}
        \caption{$\mu_{2}$ (birth).}
\end{subfigure}

\begin{subfigure}[t]{0.5\textwidth}
        \centering
        \includegraphics[width=0.90\textwidth]{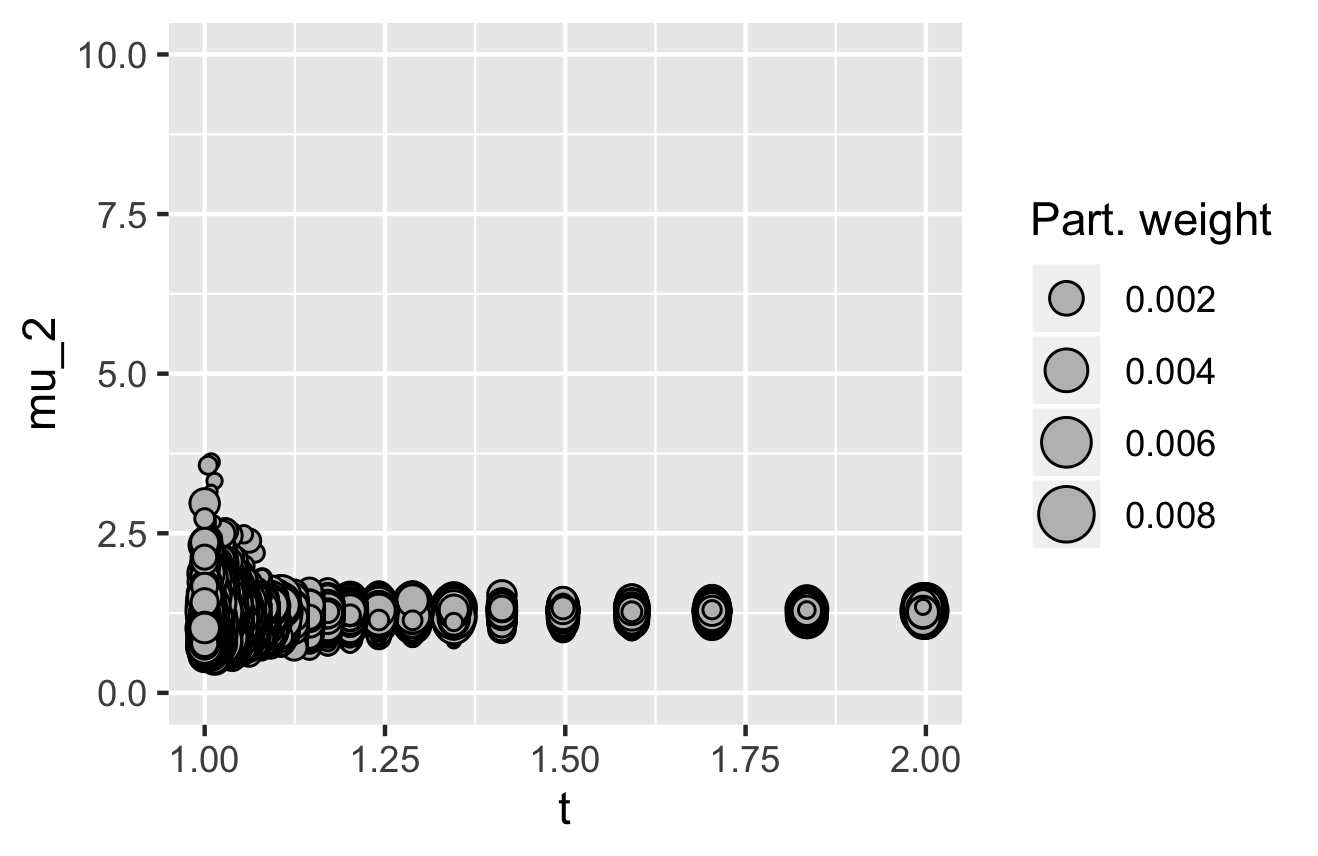}
        \caption{$\mu_{2}$ (split).}
\end{subfigure}

\\

\begin{subfigure}[t]{0.5\textwidth}
        \centering
        \includegraphics[width=0.90\textwidth]{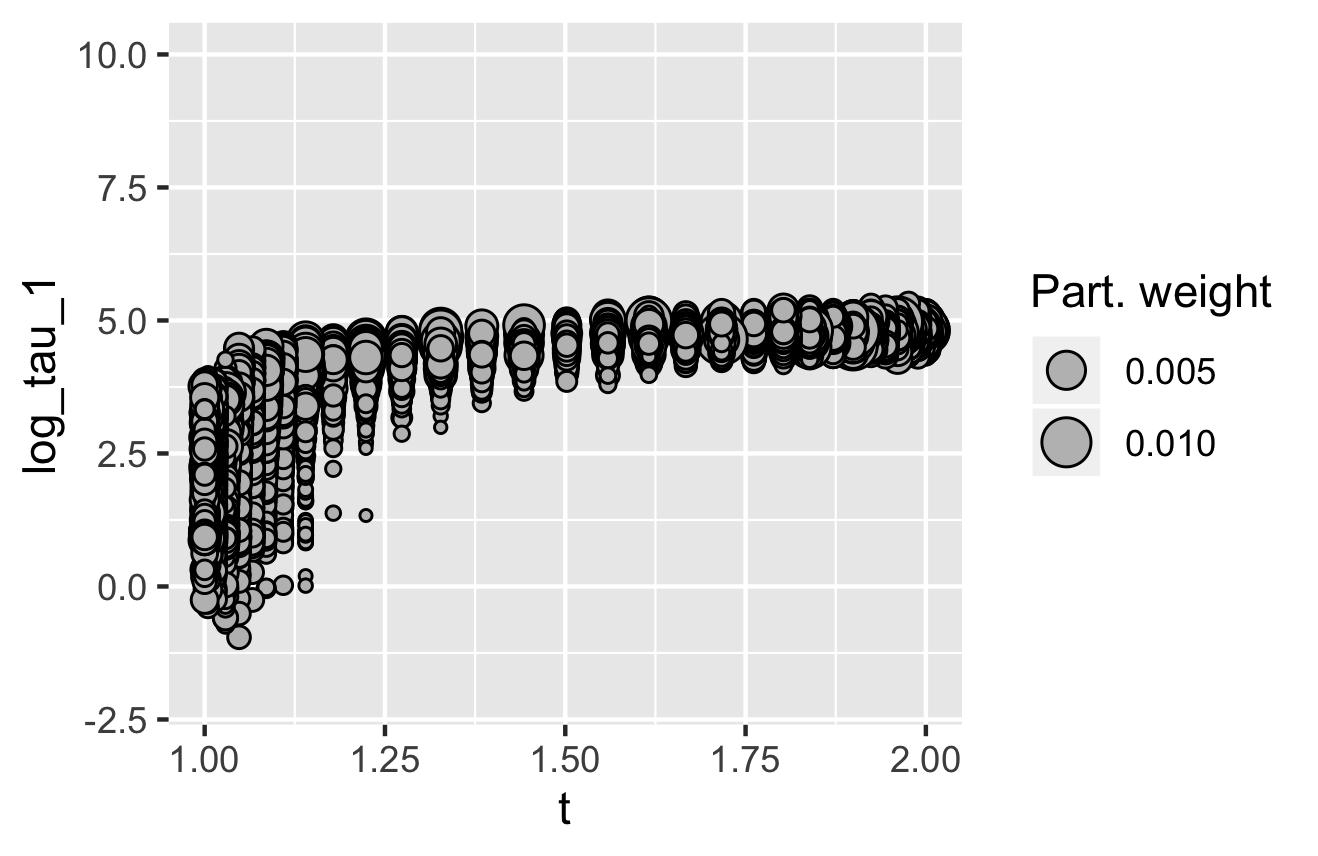}
        \caption{$\log\left(\tau_{1}\right)$ (birth).}
        \label{fig:The-evolution-of-3}
\end{subfigure}

\begin{subfigure}[t]{0.5\textwidth}
        \centering
        \includegraphics[width=0.90\textwidth]{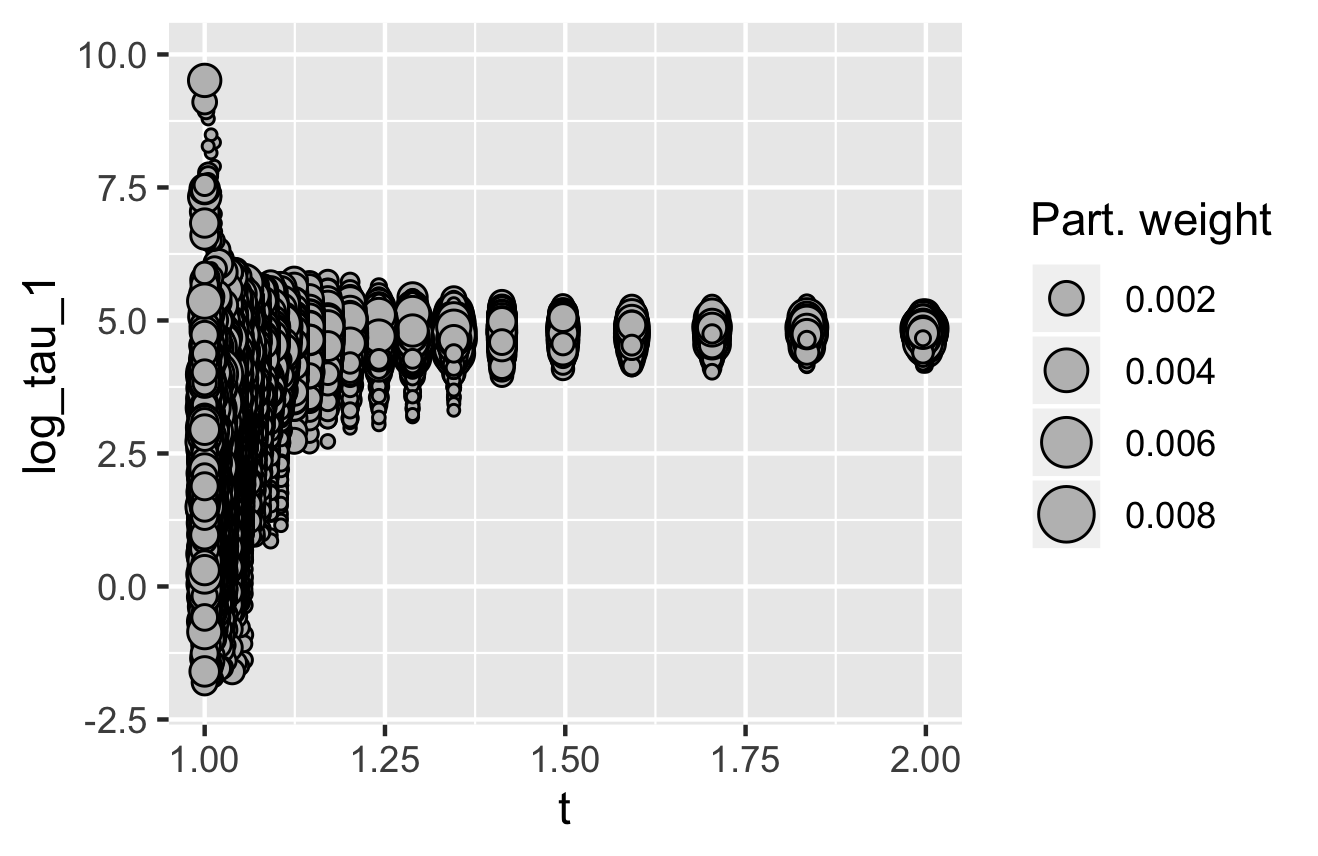}
        \caption{$\log\left(\tau_{1}\right)$ (split).}
\end{subfigure}

\\

\begin{subfigure}[t]{0.5\textwidth}
        \centering
        \includegraphics[width=0.90\textwidth]{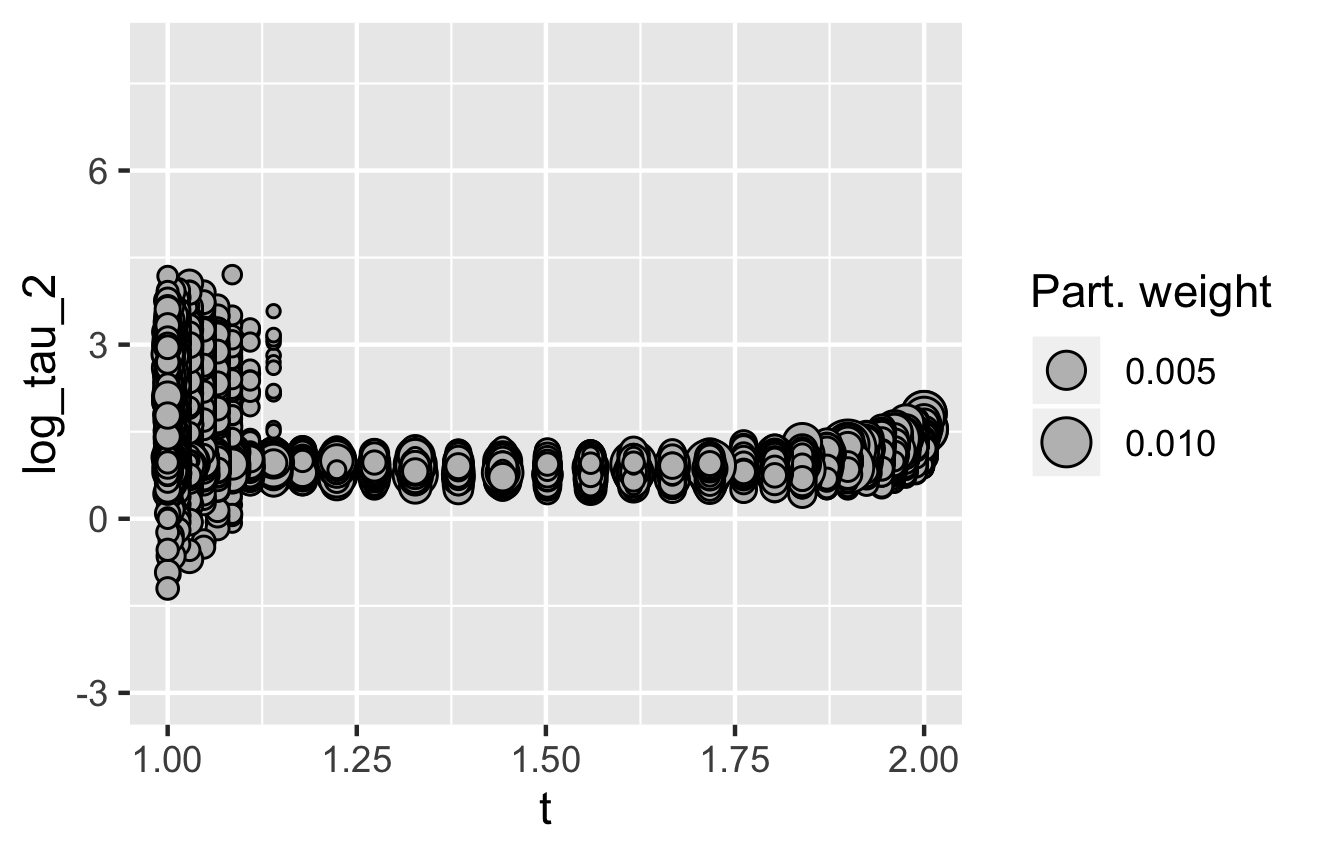}
        \caption{$\log\left(\tau_{2}\right)$ (birth).}
\end{subfigure}

\begin{subfigure}[t]{0.5\textwidth}
        \centering
        \includegraphics[width=0.90\textwidth]{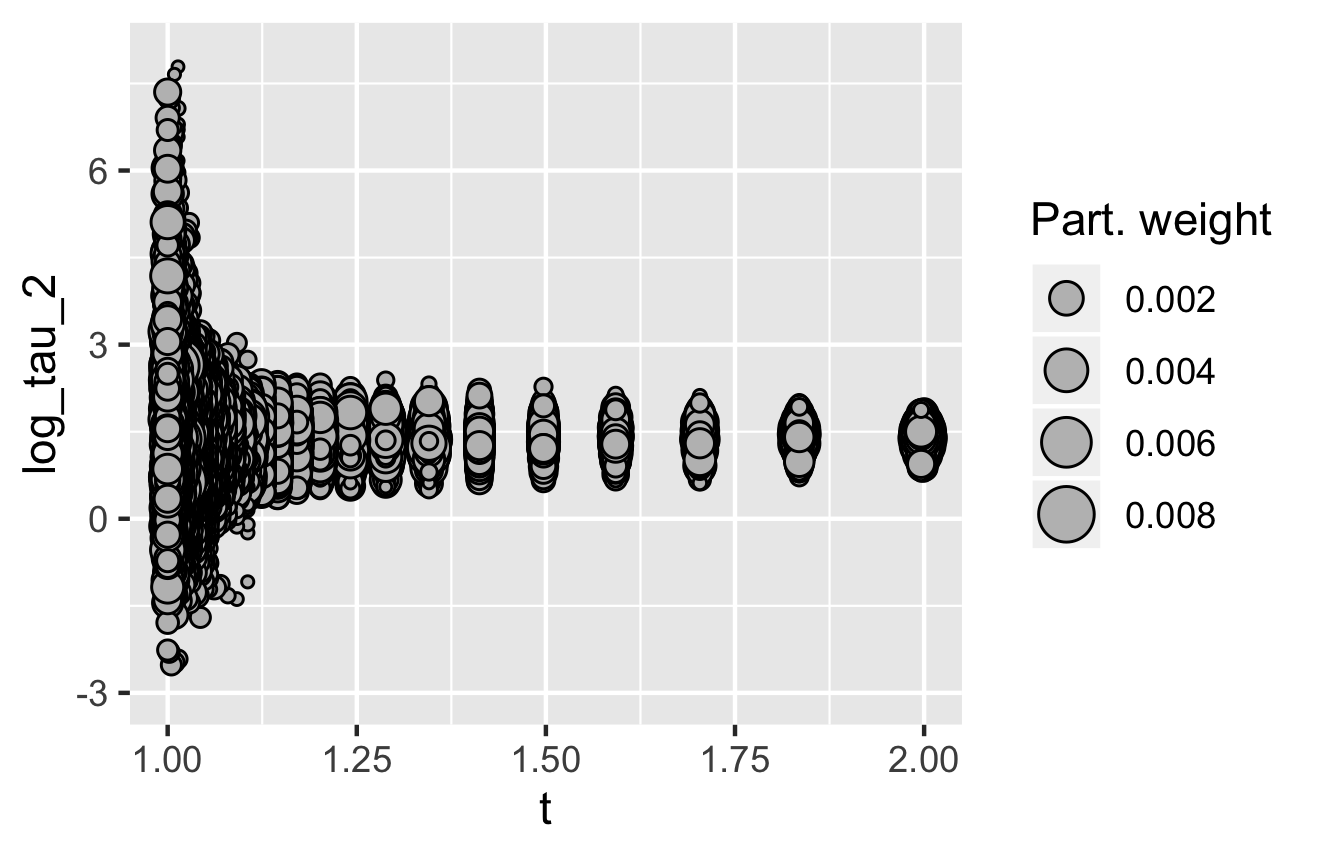}
        \caption{$\log\left(\tau_{2}\right)$ (split).}
\end{subfigure}

\end{tabular}

\caption{The evolution of particles from model 1 to model 2 for the birth and split moves on the enzyme data.}
\label{fig:Results-of-TSMC}
\end{figure*}

We ran SMC2 and the TSMC approaches on the enzyme data from \citet{Richardson1997}. We ran the algorithms 50 times,
up to a maximum of $T=8$ components, with $P=500$ particles. We used an
adaptive sequence of intermediate distributions, choosing the next
intermediate distribution to be the one that yields a CESS (equation
\ref{eq:cess}) of $\beta P$, where $\beta=0.99$. We resampled using
stratified resampling when the ESS falls below $\alpha P$, where
$\alpha=0.5$. Figure \ref{fig:Results-of-TSMC} compares the birth
and split TSMC algorithms when moving from one to two components.
We observe that the split transformation has the effect of moving
the parameters to initial values that are more appropriate for exploring
the posterior on two components. For this dataset, the birth move is a poor choice for the existing parameters in the model: figure \ref{fig:The-evolution-of-3}
shows that no particles drawn from the proposal (i.e. the posterior
for the single component model) overlap with the posterior for the
first component in the two component model. Despite the poor proposal,
the intermediate distributions (of which there are many more than
used for the split move) enable a good representation of the posterior
distribution, although below we see that the poor proposal results
in very poor estimates of the marginal likelihood.

\begin{figure}
\begin{subfigure}[t]{0.5\textwidth}
        \centering
        \includegraphics[width=0.9\textwidth]{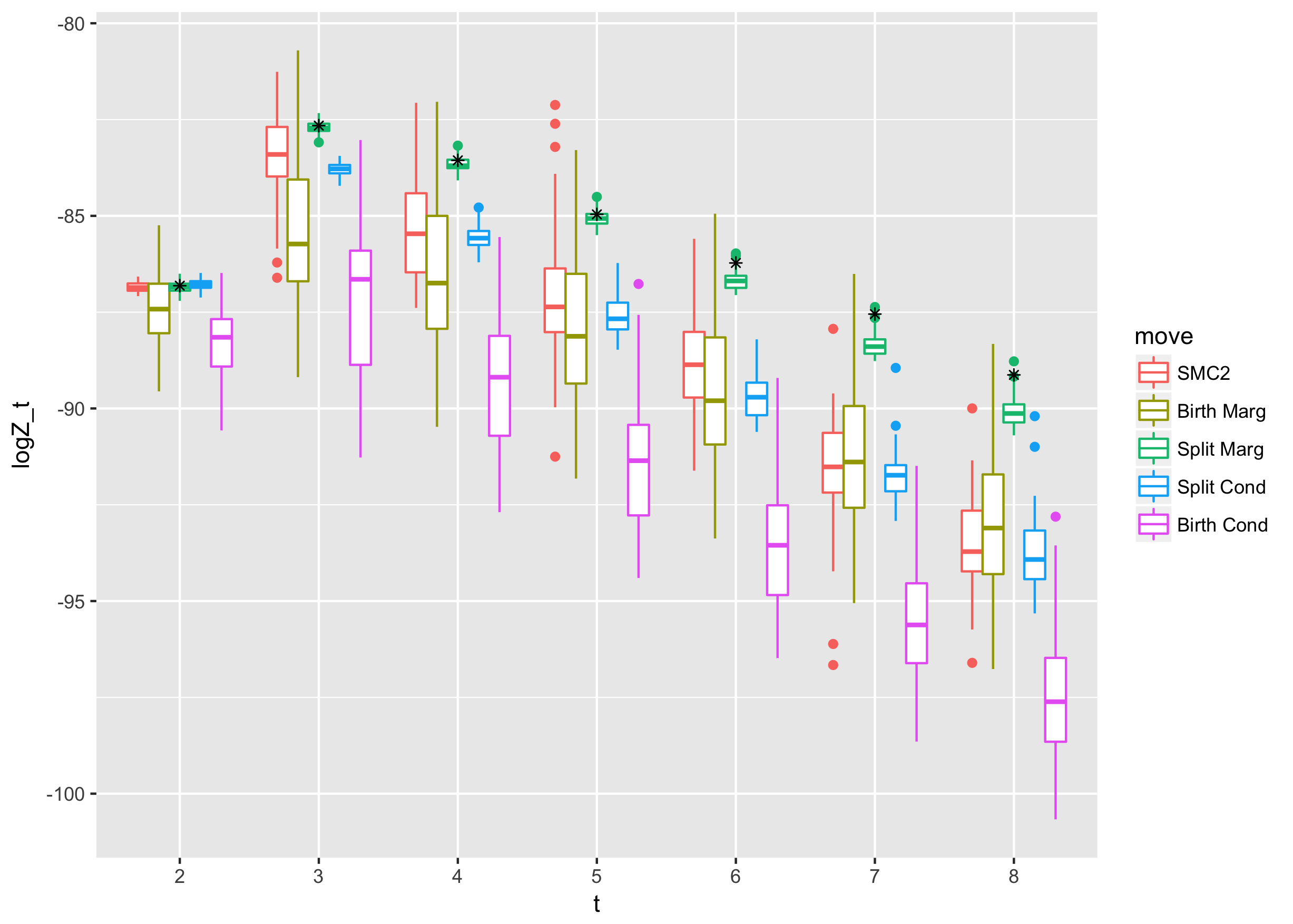}
        \caption{Box plots of the log marginal likelihood estimates from each algorithm. Black dots represent the ``truth'' computed using a long SMC2 run.}
        \label{fig:Box-plots-of}
\end{subfigure}

\begin{subfigure}[t]{0.5\textwidth}
        \centering
        \includegraphics[width=0.9\textwidth]{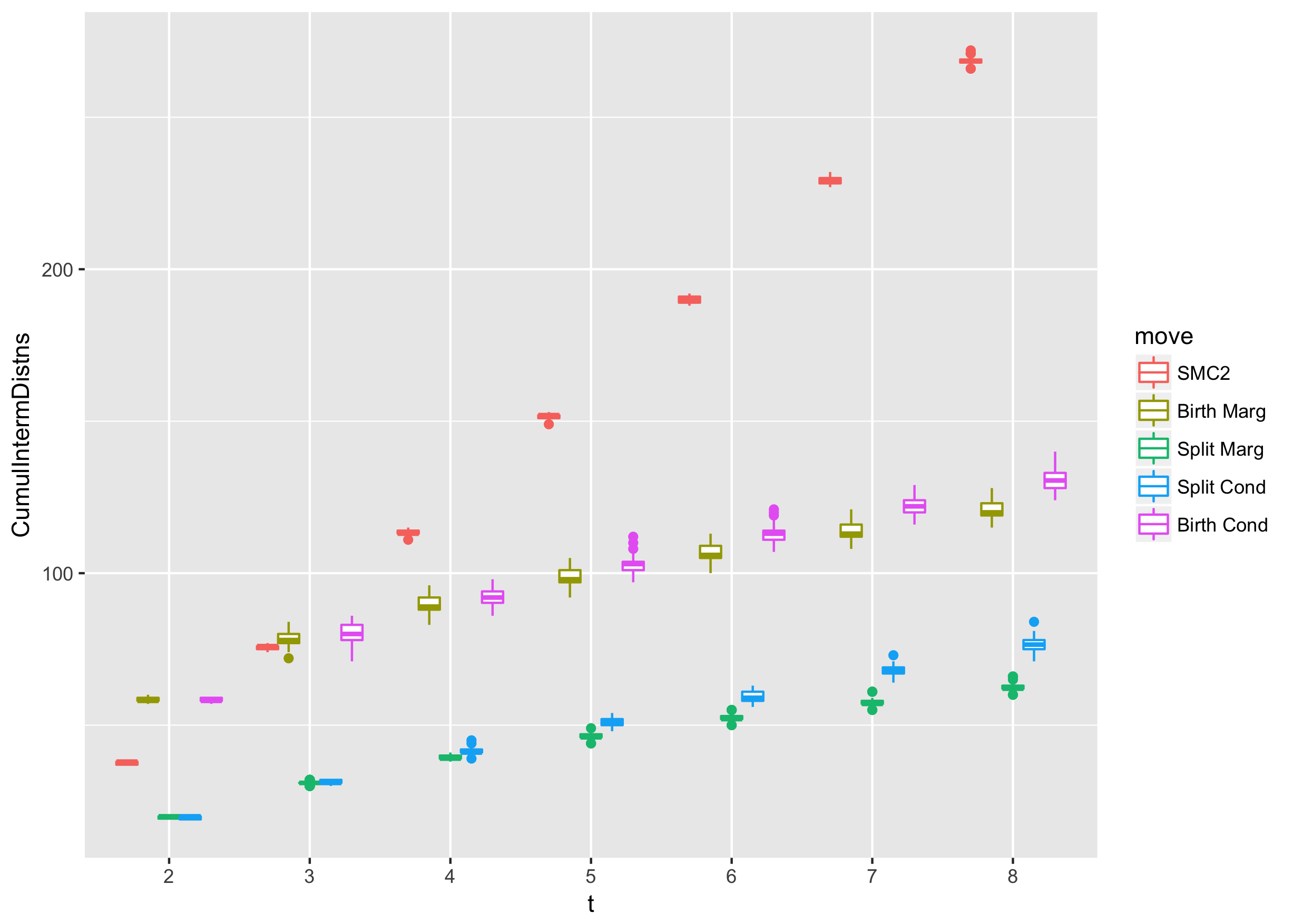}
        \caption{The cumulative number of intermediate distributions up to model t.}
        \label{fig:The-number-of}
\end{subfigure}

\begin{subfigure}[t]{0.5\textwidth}
        \centering
        \includegraphics[width=0.90\textwidth]{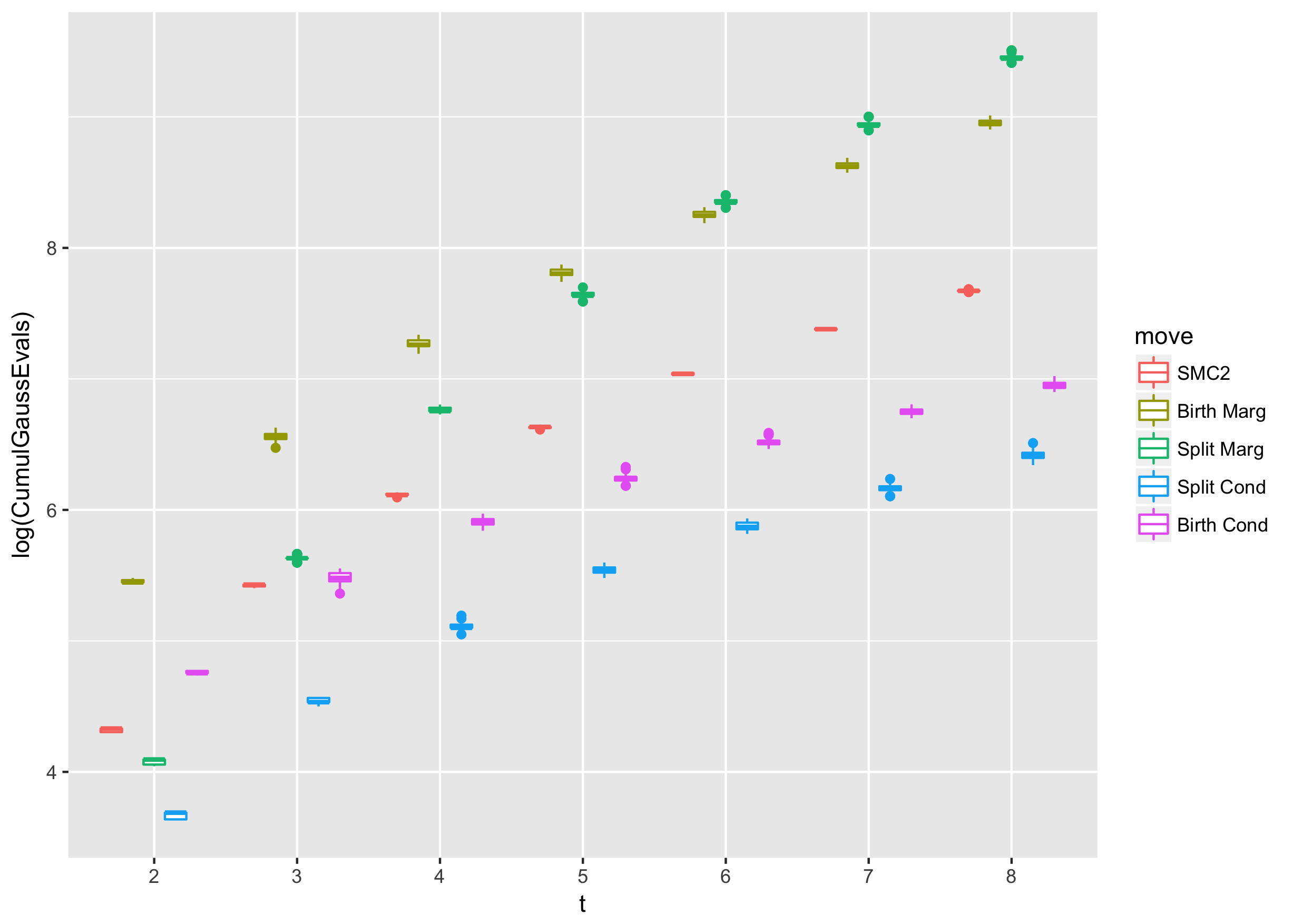}
        \caption{The cumulative number of Gaussian evaluations needed for computing the incremental weights up to model t.}
        \label{fig:GaussEvals}
\end{subfigure}

\caption{The relative performance of the different SMC schemes on the mixture example.}
\label{fig:The-relative-performance}
\end{figure}

Figure \ref{fig:Box-plots-of} shows log marginal likelihood
estimates from the different approaches (note that a poor quality
SMC usually results in an underestimate of the log marginal likelihood), and the cumulative number of intermediate distributions used in estimating all of the marginal likelihoods up to model $t$ for each $t\in\{ 1, \dots,T \}$. 
We observe that the performance of SMC2 degrades as the dimension
increases due to the increasing distance of the prior from the posterior:
we see that the adaptive scheme using the CESS results in the number
of intermediate distributions across all dimensions being approximately
constant which, as suggested by \citet{Beskos2014b} is insufficient
to control the variance as the dimension grows. As discussed above,
both birth TSMC methods yield inaccurate Bayes' factor estimates,
with split TSMC exhibiting substantially better performance. However,
we see that neither conditional approach yields very accurate results
when using the weight update given in equation (\ref{eq:tsmc_weight_route});
instead the marginalised weight update is required to provide good
estimates. The marginal version of split TSMC significantly outperforms
the other approaches, although we note that this is achieved at a
higher computational cost due to the sum in the denominator of the
weight updates, this can be observed in Figure \ref{fig:GaussEvals} which shows the cumulative number of Gaussian evaluations for computing the weights in each case. For all TSMC approaches, we see that the number of intermediate
distributions (Figure \ref{fig:The-number-of}) decreases as we increase dimension. This result can
be attributed to the relatively small change that results from only
adding a single component to the model at a time in TSMC. If the method
has a good representation of the target at model $t$ and there is
minimal change in the posterior on the existing $t$ components when
moving to model $t+1$, then the SMC is effectively only exploring
the posterior on the additional component and thus has higher ESS.

In the Supplementary Information we provide similar results for two other datasets, stressing that sensible transformations and efficient MCMC moves are essential for obtaining good estimates of the normalising constants. Interestingly, and in contrast to the enzyme data presented above, for one of these other datasets neither the split nor the birth moves outperformed SMC2; this is due to the specific distribution of the observations in such dataset.

\section{Sequential Bayesian inference under the coalescent\label{sec:Sequential-Bayesian-inference}}

\subsection{Introduction}

In this section we describe the use of TSMC for online inference under
the coalescent model in population genetics \citep{Kingman1982};
we consider the case in which we wish to infer the \emph{clonal ancestry}
(or \emph{ancestral tree}) of a bacterial population from DNA sequence
data. Current approaches in this area use MCMC \citep{Drummond2007}, which is a limitation
in situations where DNA sequence data does not arrive as a batch,
such as may happen when studying the spread of an infectious disease
as the outbreak is progressing \citep{Didelot2014}. We instead introduce
an SMC approach to online inference, inferring posterior distribution
as sequences become available (this approach is similar to that of
\citet{Dinh2016} which was devised simultaneously to ours).
We further envisage that TSMC will be useful in cases in
which data is available as a single batch, through exploiting the well known
property that a tree estimated from $t+1$ sequences is usually similar
to a tree estimated from $t$ sequences. Exploring
the space of trees for a large number of sequences appears challenging
due to the large number of possible trees: through adding leaves one
by one the SMC approach follows a path through tree space in which
transitions from distribution $\pi_{t}$ to $\pi_{t+1}$ are not challenging.
Further, our approach yields more stable estimates of the marginal
likelihood of models than current approaches used routinely in population
genetics, such as the infinite variance harmonic mean estimator \citep{Drummond2007} and the stepping stone algorithm \citep{Drummond2007,Xie2011}.

\subsubsection{Previous work}

The idea of updating a tree by adding leaves dates back to at least
\citet{Felsenstein1981a}, in which he describes, for maximum likelihood
estimation, that an effective search strategy in tree space is to
add species one by one. More recent work also makes use of the idea
of adding sequences one at a time: ARGWeaver \citep{Rasmussen2014a}
uses this approach to initialise MCMC on (in this case, a space of
graphs), $t+1$ sequences using the output of MCMC on $t$ sequences,
and TreeMix \citep{Pickrell2012} uses a similar idea in a greedy
algorithm. In work conducted simultaneously to our own, \citet{Dinh2016}
also propose a sequential Monte Carlo approach to inferring phylogenies
in which the sequence of distributions is given by introducing sequences
one by one. However, their approach: uses different proposal distributions for new sequences; does not infer the mutation rate simultaneously with the tree; does not exploit intermediate distributions to reduce the variance; and does not use adaptive MCMC moves. Further investigation of their approach can be found in  \citet{Fourment2018}, where different guided proposal distributions are explored but that still presents the aforementioned limitations.

\subsubsection{Data and model\label{subsec:Data-and-model}}

We consider the analysis of $T$ aligned genome sequences $y=y_{1:T}$,
each of length $N$. Sites that differ across sequences are known as single nucleotide polymorphisms (SNPs). The
data (which is freely available from \\ \texttt{http://pubmlst.org/saureus/}) used in our examples consists of seven ``multi-locus sequence
type'' (MLST) genes of 25 \emph{Staphylococcus aureus} sequences,
which have been chosen to provide a sample representing the worldwide
diversity of this species \citep{Everitt2014}. We make the assumption
that the population has had a constant size over time, that it evolves
clonally and that SNPs are the result of mutation. Our task is to
infer the clonal ancestry of the individuals in the study, i.e. the
tree describing how the individuals in the sample evolved from their
common ancestors, and (additional to \citet{Dinh2016}) the rate of
mutation in the population. We describe a TSMC algorithm for addressing
this problem in section \ref{subsec:TSMC-for-the}, before presenting
results in section \ref{subsec:Results-1}. In the remainder of this
section we introduce a little notation.

Let $\mathcal{T}_{t}$ represent the clonal ancestry of $t$ individuals
and let $\theta/2$ be the expected number of mutations in a generation.
We are interested in the sequence of distributions 
\[
\pi_{t}\left(\mathcal{T}_{t},\theta\mid y_{1:t}\right)\propto f\left(y_{1:t}\mid\mathcal{T}_{t},\theta\right)p\left(\mathcal{T}_{t}\right)p\left(\theta\right)
\]
for $t=1:T$. We here we use the coalescent prior \citep{Kingman1982} $p\left(\mathcal{T}_{t}\right)$
for the ancestral tree, the Jukes-Cantor substitution model \citep{Jukes1969} for
$f\left(y_{1:t}\mid\mathcal{T}_{t},\theta\right)$ and choose $p\left(\theta\right)$
to be a gamma distribution with shape 1 and rate 5 (that has its mass on biologically plausible
values of $\theta$). Let $l_{t}^{(a)}$ denote the length of time
for which $a$ branches exist in the tree, for $2\leq a\leq t$. The
heights of the coalescent events are given by $h^{(a)}=\sum_{\iota=a}^{t}l_{t}^{(\iota)}$,
with $h_{t}^{(a)}$ being the $\left(t-a+1\right)$th coalesence time
when indexing from the the leaves of the tree. We let $\mathcal{T}_{t}$
be a random vector $\left(\mathcal{B}_{t},h_{t}^{(2)},...,h_{t}^{(t)}\right)$
where $\mathcal{B}_{t}$ is itself a vector of discrete variables
representing the branching order. When we refer to a lineage of a
leaf node, this refers to the sequence of branches from this leaf
node to the root of the tree. 

\subsection{TSMC for the coalescent\label{subsec:TSMC-for-the}}

In this section we describe an approach to adding a new leaf to
an existing tree, using a transformation as in section \ref{subsec:Using-transformations-in}.
The basic idea is to first propose a lineage to add the new branch to (from distribution $\chi_{t}^{(g)}$), followed by a height $h_{\text{t}}^{(\text{new})}$ conditional
on this lineage (from distribution $\chi_{t}^{(h)}$) at which the branch connected to the new leaf will
join the tree. The resultant weight update is
\begin{align}
&\tilde{w}_{t+1}=w_{t} \frac{\pi_{t+1}\left(\mathcal{T}_{t+1},\theta\mid y_{1:t+1}\right)}{\pi_{t}\left(\mathcal{T}_{t},\theta\mid y_{1:t}\right)} \nonumber\\
&\quad \Bigg/ \Bigg(\sum_{s\in\Lambda} \Big[ \chi_{t}^{(g)}\left(g_{t}=s\mid\theta_{t},\mathcal{T}_{t},y_{1:t+1}\right) \times \chi_{t}^{(h)} \left(h_{t}^{(\text{new})}\mid g_{t}=s,\theta_{t},\mathcal{T}_{t},y_{1:t+1} \right) \Big] \Bigg)
\label{eq:lineage_then_height-weight}
\end{align}
where $\Lambda$ is the set that contains the leaves of the lineages
that if proposed, could have resulted in the new branch (under the inverse image of the transformation). Note the relationship
with equation (\ref{eq:tsmc_weight_route_rb}): we achieve a lower
variance through summing over the possible lineages rather than using
an SMC over the joint space that includes the lineage variable.

To choose the lineage, we make use of an approximation to the probability
that the new sequence is $M_{s}$ mutations from each of the existing
leaves, via approximating the pairwise likelihood of the new sequence and each existing leaf. Following \citet{Stephens2000c} (see also \citet{Li2003}) we set the probability
of choosing the lineage with leaf $s$ using
\begin{equation}
\chi_{t}^{(g)}\left(s\mid\theta_{t},y_{1:t+1}\right)\propto\left(\frac{N\theta_{t}}{t+N\theta_{t}}\right)^{M_{s}}.\label{eq:prob_the_same_coal}
\end{equation}

For $\chi_{t}^{(h)}$ we propose to approximate the pairwise likelihood
$f_{t+1,s}\left(y_{s},y_{t+1}\mid\theta,h_{t}^{(\text{new})},g_{t}=s\right)$,
where $y_{s}$ is the sequence at the leaf of the chosen lineage.
Since only two sequences are involved in this likelihood, it is likely
to have heavier tails than the posterior. We use a Laplace approximation on a transformed space, following \citet{Reis2011}: further details are given in the Supplementary Information, section 3.2.

\subsection{Results\label{subsec:Results-1}}

We used $P=250$ particles, with an adaptive sequence of intermediate distributions, choosing the next intermediate distribution to be the
one the yields a CESS (equation \ref{eq:cess}) of $\beta P$, where
$\beta=0.95$. Resampling is performed whenever the ESS falls below
$\alpha P$, where $\alpha=0.5$. At each iteration we used the current
population of particles to tune the proposal variances, as detailed in the Supplementary Information, section 3.3.

We used six different configurations
of our approach, for two different orderings of the 25 sequences.
The two orderings were chosen as follows: the ``nearest''/``furthest''
ordering was chosen by starting with the two sequences with the smallest/largest
pairwise SNP difference, then add sequences in the order of minimum/maximum
SNP difference to an existing sequence. The six configurations of
the methods were: the default configuration; using no tree topology
changing MCMC moves; taking $\chi_{t}^{(h)}$ to be an $\text{Exp}(1)$
distribution (less concentrated than the Laplace-based proposal); raising equation (\ref{eq:prob_the_same_coal}) to the
power 0 to give a uniform lineage proposal; raising equation (\ref{eq:prob_the_same_coal})
$\chi_{t}^{(g)}$ to the power 2; and raising equation (\ref{eq:prob_the_same_coal})
$\chi_{t}^{(g)}$ to the power 4. These latter two approaches use
a lineage proposal where the probability is more concentrated on a
smaller number of lineages.

\begin{figure*}

\begin{tabular}{cc}

\begin{subfigure}[t]{0.5\textwidth}
        \centering
        \includegraphics[width=1.2\textwidth]{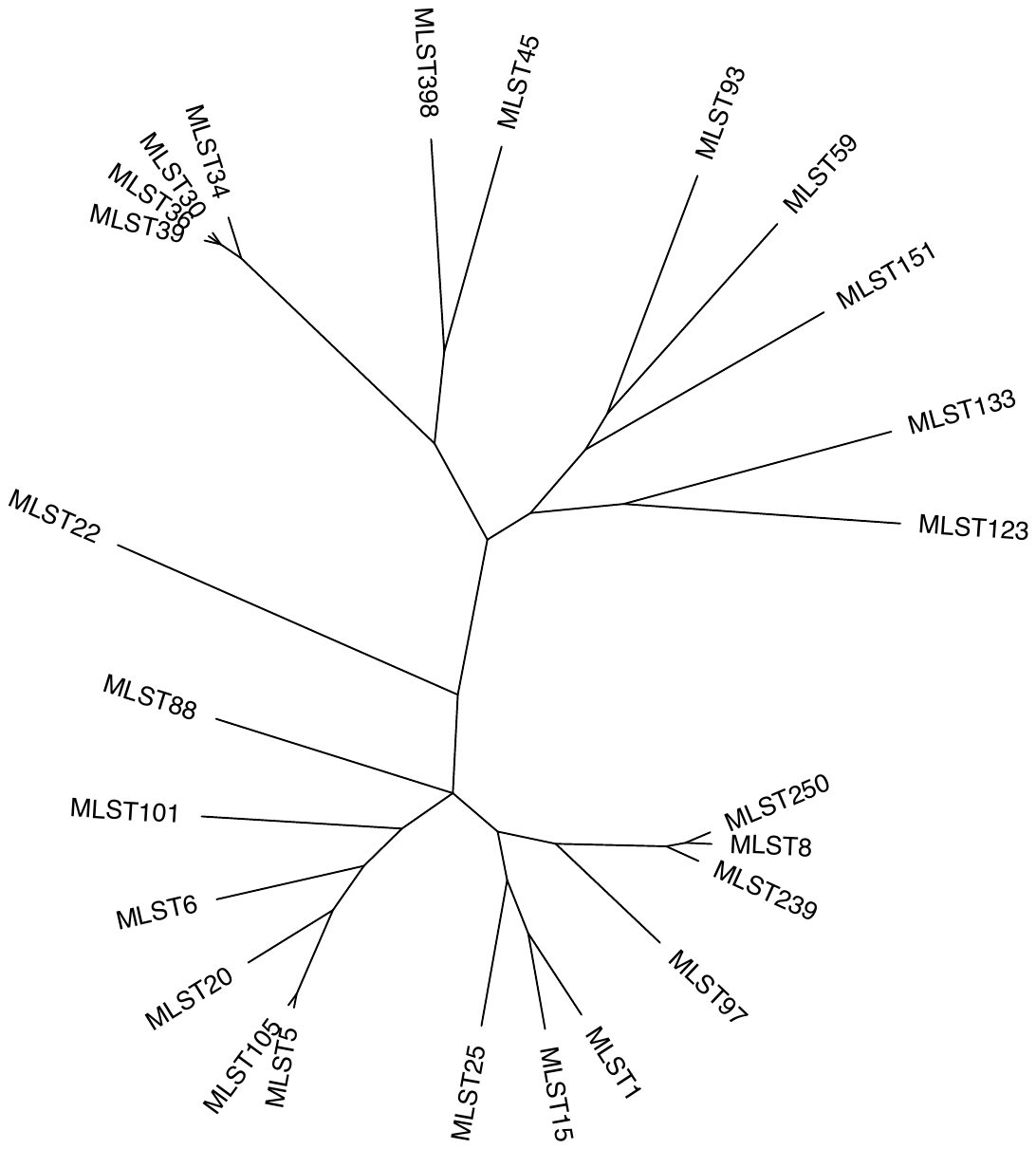}
        \caption{MCMC using 50 million iterations.}
        \label{fig:Default-configuration-MCMC}
\end{subfigure}

\begin{subfigure}[t]{0.5\textwidth}
        \centering
        \includegraphics[width=1.2\textwidth]{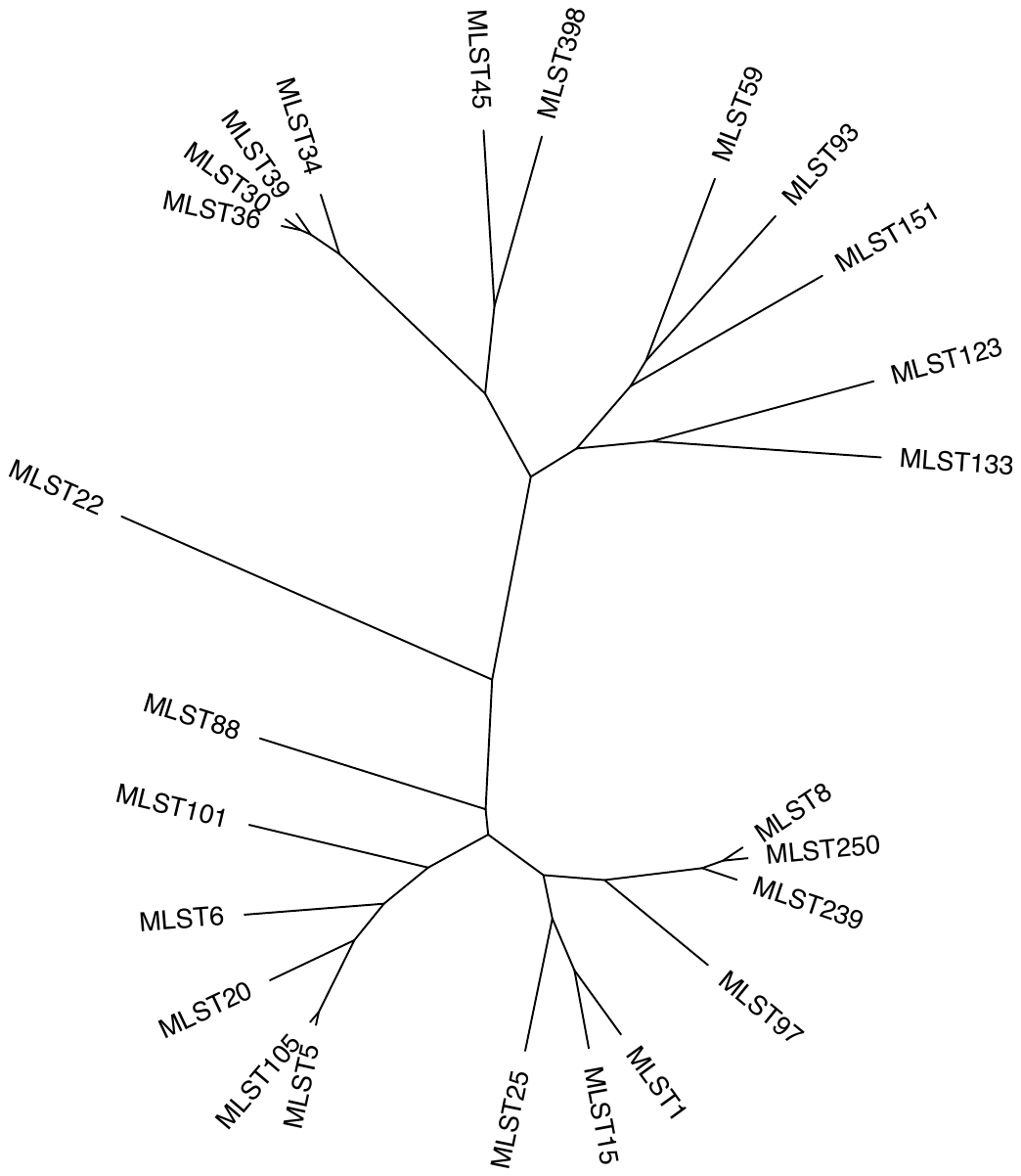}
        \caption{Default configuration using ``nearest'' ordering.}
        \label{fig:Default-configuration-of}
\end{subfigure}

\\

\begin{subfigure}[t]{0.5\textwidth}
        \centering
        \includegraphics[width=1.2\textwidth]{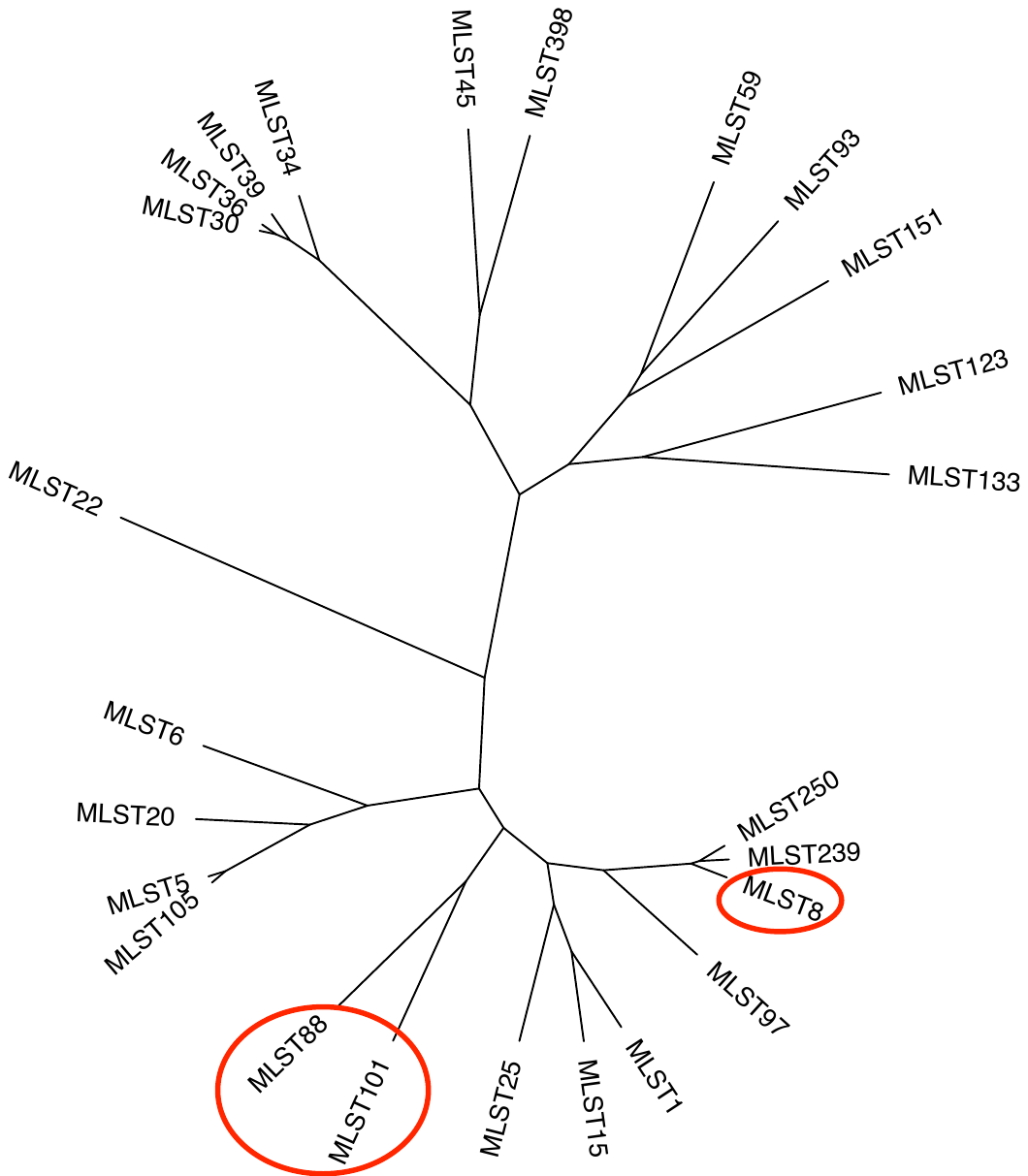}
        \caption{No topology moves using ``furthest'' ordering.}
        \label{fig:TSMC-with-no}
\end{subfigure}

\begin{subfigure}[t]{0.5\textwidth}
        \centering
        \includegraphics[width=1.2\textwidth]{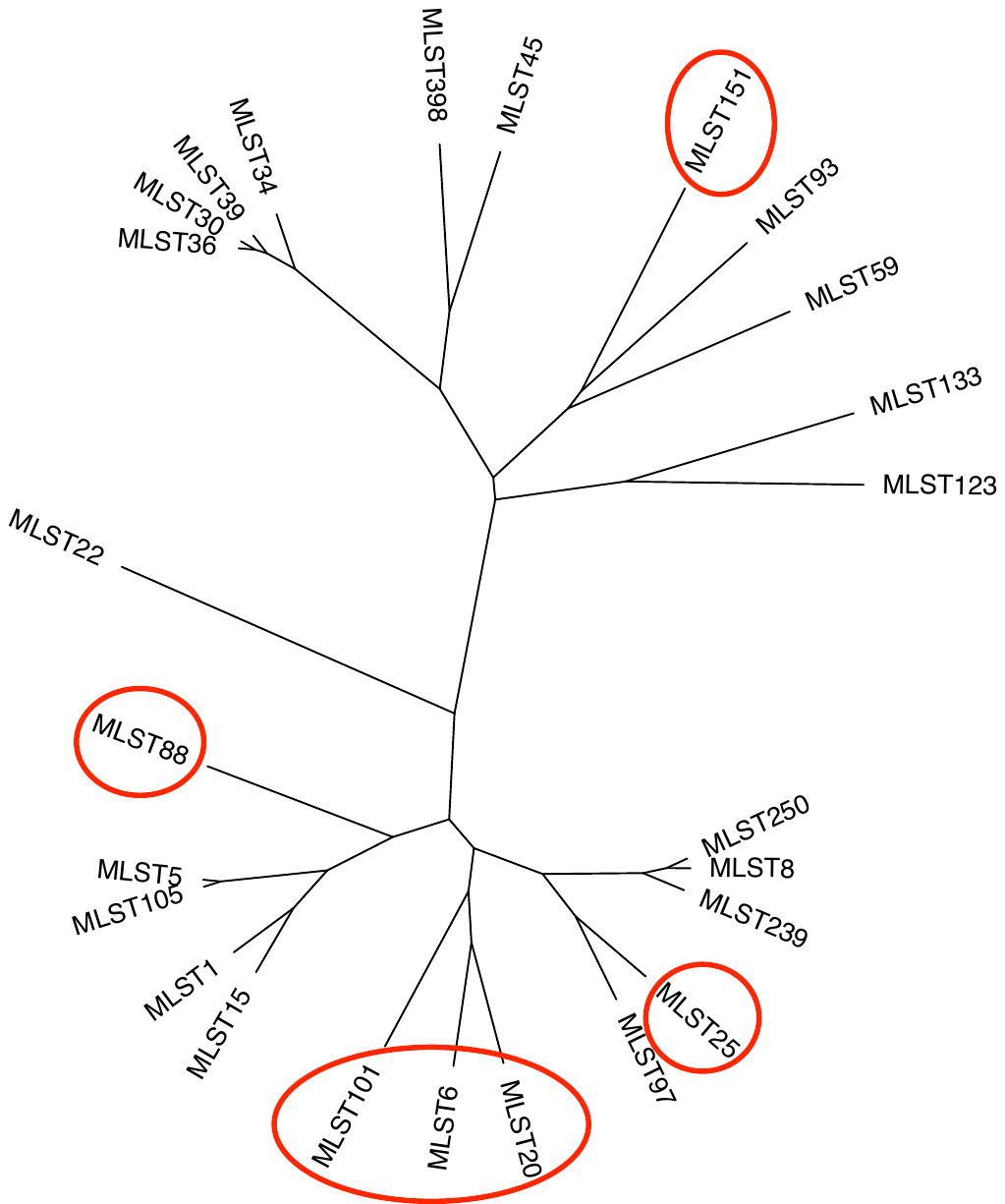}
        \caption{No topology moves using ``nearest'' ordering.}
        \label{fig:TSMC-with-no-1}
\end{subfigure}

\end{tabular}
\caption{Majority-rule consensus trees found by MCMC and the default configuration of TSMC (top), and different configurations of TSMC (bottom) with differences to the result obtained by the default configuration highlighted.}
\label{fig:Majority-rule-consensus-trees}
\end{figure*}

Figure \ref{fig:Majority-rule-consensus-trees} shows majority-rule
consensus trees from an MCMC run and the final TSMC iterations. Figure \ref{fig:Default-configuration-of}
is generated by the default configuration (for the ``furthest'' ordering,
although results from the ``nearest'' ordering are nearly identical),
and is close to the ground truth in Figure \ref{fig:Default-configuration-MCMC} (as determined by a long MCMC run).
Figures \ref{fig:TSMC-with-no} and \ref{fig:TSMC-with-no-1}
used no topology changing MCMC moves, thus illustrate the contribution
of the SMC proposal in determining the topology. Table \ref{tab:Log-marginal-likelihood}
shows estimates of the log marginal likelihood from each configuration
of the algorithm for both orderings (longer runs of our method suggest
the true value is $\approx-6333$), along with the total number of
intermediate distributions used. Recall that a poorer quality SMC
usually results in an underestimate of the log marginal likelihood,
and the number of intermediate distributions offers an indication
as to the distance between the target and the proposal where the proposal
has heavier tails than the target. We draw the following conclusions:
\begin{itemize}
\item As also suggested by figure \ref{fig:Majority-rule-consensus-trees},
we see that the ``furthest'' ordering provides consistently better
results than the ``nearest'' ordering. ``Furthest'' provides an
ordering in which new sequences are often added above the root of
the current tree, since the existing sequences are all more closely
related than the new sequence, whereas ``nearest'' frequently results
in adding a leaf close to the existing leaves of the tree. In the
latter strategy, the proposal relating to the new sequence is often
good, but adding a new sequence can have a large effect on the posterior
of existing variables. We see this by comparing figures (\ref{fig:TSMC-with-no})
and (\ref{fig:TSMC-with-no-1}), observing that the ``furthest''
ordering results in a topology that is close to the truth. The topology
from the ``nearest'' ordering is not as close to the truth, thus
is more reliant on topology changing MCMC moves to give an accurate
sample from the posterior.
\item As expected, using no MCMC topology moves results in very poor estimates,
highlighting the important role of MCMC in generating diversity not
introduced in the SMC proposals. This poor quality is not accounted
for by the adaptive scheme based on the CESS introducing more intermediate
distributions, since the CESS is only based on the weights of the
particles and cannot account for a lack of diversity.
\item Using less directed proposals, on both the lineage and the height,
increases the distance between the proposal and target, and results
in lower quality estimates.
\item Using more directed proposals on the lineage may in some cases slightly
improve the method, but appear to make the method less robust to the
order in which the individuals are added (so may not be suitable in
applications where the order of the individuals cannot be chosen).
\end{itemize}

\begin{table*}
\centering
\begin{tabular}{|c|c|c|c|c|c|c|}
\hline 
Default & No top. moves & $\chi_{t}^{(h)}=\text{Exp}(1)$ & $\left(\chi_{t}^{(h)}\right)^{0}$ & $\left(\chi_{t}^{(h)}\right)^{2}$ & $\left(\chi_{t}^{(h)}\right)^{4}$\tabularnewline
\hline 
\hline 
-6333.9 / 267 & -6338.8 / 257 & -6335.1 / 408 & -6336.9 / 330 & -6333.1 / 247 & -6334.3 / 238\tabularnewline
\hline 
-6335.8 / 323 & -6354.6 / 293 & -6337.8 / 501 & -6341.0 / 384 & -6339.0 / 300 & -6342.0 / 255\tabularnewline
\hline 
\end{tabular}

\caption{Log marginal likelihood estimates and total number of distributions
for TSMC applied to the coalescent (5 s.f.), for the ``Furthest'' (first line) and ``Nearest'' (second line) orderings. \label{tab:Log-marginal-likelihood}}

\end{table*}

A video showing the evolution of the majority rule consensus tree (and the marginal likelihood estimate) through all iterations of the SMC, using the default configuration, can be found at\\https://www.youtube.com/watch?v=pSDK9ajm2OY.

\section{Conclusions\label{sec:Conclusions}}

This paper introduces a sequential technique for Bayesian model comparison
and parameter estimation, and an approach to online parameter and
marginal likelihood estimation for the coalescent, underpinned by
the same methodological development: TSMC. We show that whilst TSMC
performs inference on a sequence of posterior distributions with increasing
dimension, it is a special case of the standard SMC sampler framework
of \citet{DelMoral2007}. In this section we outline several points
that are not described elsewhere.

One innovation introduced in the paper is the use of transformations
within SMC for creating proposal distributions when moving between
dimensions. The effectiveness of TSMC is governed by the distance
between neighbouring distributions, thus to design TSMC algorithms
suitable for any given application, we require the design of a suitable
transformation that minimises the distance between neighbouring distributions.
This is essentially the same challenge as is faced in designing effective
RJMCMC algorithms, and we may make use of many of the methods devised
in the RJMCMC literature \citep{Hastie2012a}. The ideal case is to
use a transformation such that every distribution $\varphi_{t\rightarrow T}$
becomes identical, in which case one may simulate from $\pi_{T}$
simply by simulating from $\pi_{0}$ then applying the transformation.
Approximating such a ``transport map'' for a sequence of continuous
distributions is described in \citet{Heng2015}. As discussed in section
\ref{subsec:Targets-on-spaces}, \citet{Heng2015} is one of a number
of papers that seeks to automatically construct useful transformations,
and we anticipate these techniques being of use in the case of changing
dimension that is addressed in this paper. In the RJMCMC literature,
\citet{Brooks2003b} describe methods for automatically constructing
the ``fill in'' distributions $\psi_{t}$ for a given transformation:
the literature on transport maps could be used to automatically construct
the transformation in advance of this step.

In figure \ref{fig:The-relative-performance} of section \ref{sec:Bayesian-model-comparison}
we see a characteristic of this approach that will be common
to many applications, in that the estimated marginal likelihood rises
as the model is improved, then falls as the effect of the model complexity
penalisation becomes more influential than improvements to the likelihood.
We note that by using estimates of the variance of the marginal likelihood
estimate \citep{Lee2016}, we may construct a formal diagnostic that
decides to terminate the algorithm at a particular model, on observing
that the estimated marginal likelihood declines from an estimated
maximum value.

Although the examples in this paper both involve posterior distributions
of increasing dimension, we also see a use for our approach in some
cases that involve a distributions of decreasing dimension. For example,
in population genetics, it is common to perform a large number of
different analyses using different overlapping sets of sequences.
For this reason many practitioners would value an inference technique
that allows for the removal, as well as the addition, of sequences. Further, many genetics applications now involve the analysis of whole genome sequences. Our approach is applicable in this setting, and for this purpose a BEAST2 package is currently under development.



\begin{acknowledgements}
Thanks to Christophe Andrieu, Adam Johansen and Changqiong Wang for useful discussions; Xavier Didelot and Dan Lawson for establishing the novelty of the approach; and Christian Robert for the suggestion to use Rao-Blackwellisation in the mixture example. First and third authors were supported by BBSRC grant BB/N00874X/1. Second author was supported by the University of Reading, and the Modernising Medical Microbiology group, NDM Experimental Medicine, University of Oxford. Fourth author is a Sir Henry Dale Fellow, jointly funded by the Wellcome Trust and the Royal Society (grant 101237/Z/13/Z).
\end{acknowledgements}

\appendix

\section{Theoretical aspects\label{subsec:Theoretical-aspects}}

We define a sequence of targets $\left(\varphi_{t\rightarrow T}\right)_{t=0}^{T}$
on the same measurable space $\left(E_{T},\mathcal{E}_{T}\right)$
as follows. Consider $\vartheta_{t}\sim\varphi_{t}\left(\cdot\right)$
and define
\begin{eqnarray*}
\varphi_{t\rightarrow T}\left(\cdot\right) & := & \mathcal{L}\left(G_{t\rightarrow T}\left(\vartheta_{t}\right)\right),
\end{eqnarray*}
where $\mathcal{L}\left(X\right)$ denotes the law of a random variable
$X$, and where $G_{t\rightarrow T}$ is defined recursively for $0\leq t<T$
as follows 
\[
G_{t\rightarrow T}:=G_{t+1\rightarrow T}\circ G_{t\rightarrow t+1},
\]
with $G_{T\rightarrow T}$ as the identity function. As before, the
distributions $\varphi_{t\rightarrow T}$, $\varphi_{t}$ and $\pi_{t}$
all share the same normalising constant $Z_{t}$. Hence, the TSMC
algorithm described in the main paper can be seen as an SMC sampler
for the targets $\left\{ \varphi_{t\rightarrow T}\right\} _{t}$,
propagating particles $\left\{ \vartheta_{t\rightarrow T}^{\left(p\right)}\right\} _{t,p}$
using a sequence of MCMC kernels $\left\{ K_{t\rightarrow T}:E_{T}\times\mathcal{E}_{T}\rightarrow\left[0,1\right]\right\} _{t}$,
where $K_{t\rightarrow T}$ admits $\varphi_{t\rightarrow T}$ as
invariant. This will also be the case for the modified TSMC algorithm
with intermediate distributions, however the details are omitted for
simplicity. 

Therefore, after the $(t+1)$th iteration the target $\varphi_{t+1\rightarrow T}$
can be approximated using
\begin{eqnarray*}
\hat{\varphi}_{t+1\rightarrow T}^{P} & = & \sum_{p=1}^{P}w_{t+1\rightarrow T}^{\left(p\right)}\delta_{\vartheta_{t+1\rightarrow T}^{\left(p\right)}},
\end{eqnarray*}
where, for every $p\in\left\{ 1,\dots,P\right\} $,
\begin{eqnarray*}
w_{t+1\rightarrow T}^{\left(p\right)} & \propto & w_{t\rightarrow T}^{\left(p\right)}\frac{\tilde{\varphi}_{t+1\rightarrow T}\left(\vartheta_{t\rightarrow T}^{\left(p\right)}\right)}{\tilde{\varphi}_{t\rightarrow T}\left(\vartheta_{t\rightarrow T}^{\left(p\right)}\right)}
\end{eqnarray*}
and $w_{0\rightarrow T}^{\left(p\right)}=1/P$. Furthermore, notice
that $w_{t\rightarrow T}^{\left(p\right)}=w_{t}^{\left(p\right)}$
for all $0\leq t<T$ and any $p\in\left\{ 1,\dots,P\right\} $, consequently
expectations of the form $\varphi_{t+1}\left(h\right)$ (for a function
$h:E_{t}\rightarrow\mathbb{R}$) can be approximated using 
\begin{eqnarray*}
\hat{\varphi}_{t+1\rightarrow T}^{P}\left(h\circ G_{T\rightarrow t+1}\right) & = & \sum_{p=1}^{P}w_{t+1\rightarrow T}^{\left(p\right)}h\circ G_{T\rightarrow t+1}\left(\vartheta_{t+1\rightarrow T}^{\left(p\right)}\right)\\
 & = & \sum_{p=1}^{P}w_{t+1}^{\left(p\right)}h\left(\vartheta_{t+1}^{\left(p\right)}\right)=\hat{\varphi}_{t+1}^{P}\left(h\right).
\end{eqnarray*}
The following theorem, the proof of which may be found in \citet[Proposition 2]{DelMoral2006c},
follows from well-known standard SMC convergence results.
\begin{thm}
Under weak integrability conditions (see \citealp[Theorem 1]{Chopin2004b}
or \citealp[p300-306]{DelMoral2004}) and for any bounded $h:E_{t}\rightarrow\mathbb{R}$
, as $P\rightarrow\infty$
\end{thm}
\begin{enumerate}
\item $P^{1/2}\left\{ \hat{\varphi}_{t}^{P}\left(h\right)-\bar{\varphi}_{t}\left(h\right)\right\} \Rightarrow\mathcal{N}\left(\cdot\left|0,\sigma_{IS,t}^{2}\left(h\right)\right.\right)$,
if no resampling is performed;
\item $P^{1/2}\left\{ \hat{\varphi}_{t}^{P}\left(h\right)-\bar{\varphi}_{t}\left(h\right)\right\} \Rightarrow\mathcal{N}\left(\cdot\left|0,\sigma_{SMC,t}^{2}\left(h\right)\right.\right)$,
when multinomial resampling is performed at every iteration;
\end{enumerate}
where $\sigma_{IS,t}^{2}\left(h\right)$ and $\sigma_{SMC,t}^{2}\left(h\right)$
follow similar expressions to those in \citet[Proposition 2]{DelMoral2006c}.
\begin{rem}
As noted also in \citet{DelMoral2006c}, under strong mixing assumptions,
the variance $\sigma_{SMC,t}^{2}\left(h\right)$ can be uniformly
bounded in $t$ whereas $\sigma_{IS,t}^{2}\left(h\right)$ will typically
diverge as $t$ increases.
\end{rem}
Respecting the normalising constants $\left\{ Z_{t}\right\} _{t=1}^{T}$,
they can be approximated using 
\begin{eqnarray*}
\hat{Z}_{t+1}^{P} & = & \prod_{s=1}^{t+1}\sum_{p=1}^{P}w_{s\rightarrow T}^{\left(p\right)}\frac{\tilde{\varphi}_{s+1\rightarrow T}\left(\vartheta_{s\rightarrow T}^{\left(p\right)}\right)}{\tilde{\varphi}_{s\rightarrow T}\left(\vartheta_{s\rightarrow T}^{\left(p\right)}\right)}=\prod_{s=1}^{t+1}\sum_{p=1}^{P}w_{s}^{\left(p\right)}\frac{\tilde{\varphi}_{s+1\rightarrow s}\left(\vartheta_{s}^{\left(p\right)}\right)}{\tilde{\varphi}_{s}\left(\vartheta_{s}^{\left(p\right)}\right)},
\end{eqnarray*}
and standard results show that these estimates are unbiased (see e.g.
\citealp[proposition 7.4.1]{DelMoral2004}), with relative variance
increasing at most linearly in $t$ \citep[Theorem 5.1]{Cerou2011}.
Such results are summarised in the following theorem. 
\begin{thm}
\label{thm:expec.varian.normconsts}For fixed $E_{T}$, and when resampling
is not done adaptively, the estimates $\left\{ \hat{Z}_{t}^{P}\right\} _{t}$
satisfy 
\begin{align*}
\mathbb{E}\left[\hat{Z}_{t}^{P}\right] & =Z_{t}.
\end{align*}
Furthermore, under strong mixing assumptions there exists a constant
$C_{T}\left(t\right)$, which is linear in $t$, such that
\begin{align*}
\mathbb{V}\left[\frac{\hat{Z}_{t}^{P}}{Z_{t}}\right] & \leq\frac{C_{T}\left(t\right)}{P}.
\end{align*}
\end{thm}
However, as $T$ increases the dimension of $E_{T}$ (denoted hereafter
by $d_{T})$ may increase and we will usually require an exponential
growth in the number of particles $P$ in order to obtain meaningful
results, see e.g. \citet{Bickel2008}. For instance, without the resampling
step the ESS at time $t+1$ is closely related to the following quantity
(see e.g. \citealp{Agapiou2015})
\begin{eqnarray*}
\rho_{t+1}\left(d_{T}\right) & := & \mathbb{E}\left[\left(\prod_{s=1}^{t+1}\frac{\varphi_{s\rightarrow T}\left(\vartheta_{s-1\rightarrow T}\right)}{\varphi_{s-1\rightarrow T}\left(\vartheta_{s-1\rightarrow T}\right)}\right)^{2}\right],
\end{eqnarray*}
which serves as a measure of the dissimilarity between proposals and
targets, and that quite often increases exponentially in $d_{T}$.
This quantity provides information about the limiting proportion of
effective number of particles since
\begin{eqnarray*}
\lim_{P\rightarrow\infty}\left(\frac{\mbox{ESS}_{t+1}^{P}}{P}\right)^{-1} & = & \lim_{P\rightarrow\infty}P\sum_{p=1}^{P}\left(w_{t+1\rightarrow T}^{(p)}\right)^{2}=\lim_{P\rightarrow\infty}\frac{\frac{1}{P}\sum_{p=1}^{P}\left(w_{0}^{\left(p\right)}\prod_{s=1}^{t+1}\frac{\tilde{\varphi}_{s\rightarrow T}\left(\vartheta_{s-1\rightarrow T}^{\left(p\right)}\right)}{\tilde{\varphi}_{s-1\rightarrow T}\left(\vartheta_{s-1\rightarrow T}^{\left(p\right)}\right)}\right)^{2}}{\left(\frac{1}{P}\sum_{p=1}^{P}w_{0}^{\left(p\right)}\prod_{s=1}^{t+1}\frac{\tilde{\varphi}_{s\rightarrow T}\left(\vartheta_{s-1\rightarrow T}^{\left(p\right)}\right)}{\tilde{\varphi}_{s-1\rightarrow T}\left(\vartheta_{s-1\rightarrow T}^{\left(p\right)}\right)}\right)^{2}}\\
 & = & \frac{\mathbb{E}\left[\left(\prod_{s=1}^{t+1}\frac{\tilde{\varphi}_{s\rightarrow T}\left(\vartheta_{s-1\rightarrow T}\right)}{\tilde{\varphi}_{s-1\rightarrow T}\left(\vartheta_{s-1\rightarrow T}\right)}\right)^{2}\right]}{\left(\mathbb{E}\left[\prod_{s=1}^{t+1}\frac{\tilde{\varphi}_{s\rightarrow T}\left(\vartheta_{s-1\rightarrow T}\right)}{\tilde{\varphi}_{s-1\rightarrow T}\left(\vartheta_{s-1\rightarrow T}\right)}\right]\right)^{2}}=\rho_{t+1}.
\end{eqnarray*}
The above equation implies that $P=\mathcal{O}\left(\rho_{t+1}\left(d_{T}\right)\right)$
if we want to maintain an acceptable level for the ESS. In our context,
even though the targets $\left(\bar{\varphi}_{s\rightarrow T}\right)_{s}$
are $d_{T}$-dimensional the ratios of densities $\left(\varphi_{s\rightarrow T}/\varphi_{s-1\rightarrow T}\right)_{s}$
will involve cancellations of ``fill in'' variables as discussed
in the paper. This potentially leads to a much lower effective dimension
of the problem than $d_{T}$. 

For the SMC method presented in \citet{Dinh2016} in the context of
phylogenetic trees, the authors have shown that $\rho_{T}$ grows
at most linearly in $T$ under some strong conditions, somewhat comparable
to the strong mixing conditions required in Theorem \ref{thm:expec.varian.normconsts}.
Imposing an extra condition on the average branch length of the tree,
$\rho_{T}$ can be bounded uniformly in $T$. However, their method
performs MH moves after resampling for improving the diversity of
the particles, which could result in a sub-optimal algorithm. In contrast,
TSMC uses MH moves for bridging $\varphi_{t}$ and $\varphi_{t+1}$
via the sequence of intermediate distributions $\left(\varphi_{t,k}\right)_{k=1}^{K}$.
Heuristically, the introduction of these intermediate distributions
together with sensible transformations $\left\{ G_{t\rightarrow t+1}\right\} $
should alleviate problems due to the dissimilarity of targets, thus
providing control over $\rho_{T}$.

In this respect, the authors in \citet{Beskos2014b} have analysed
the stability of SMC samplers as the dimension of the state-space
increases when the number of particles $P$ is fixed. Their work provides
justification, to some extent, for the use of intermediate distributions
$\left(\varphi_{t,k}\right)_{k=1}^{K}$. Under some assumptions, it
has been shown that when the number of intermediate distributions
$K=\mathcal{O}\left(d_{T}\right)$, and as $d_{T}\rightarrow\infty$,
the effective sample size $\mbox{ESS}_{t+1}^{P}$ is stable in the
sense that it converges to a non-trivial random variable taking values
in $\left(1,P\right)$. The total computational cost for bridging
$\varphi_{t}$ and $\varphi_{t+1}$, assuming a product form of $d_{T}$
components, is $\mathcal{O}\left(Pd_{T}^{2}\right)$. Using this reasoning,
we suspect TSMC will work well in similar and more complex scenarios,
e.g. when the targets do not follow a product form or when strong
mixing assumptions do not hold. This idea is supported by the results
described in the paper.

\section{Bayesian model comparison for mixtures of Gaussians\label{sec:Bayesian-model-comparison}}

\subsection{Split move\label{subsec:Split-TSMC}}

Suppose that at time $t$ the transformation $G_{t\rightarrow t+1}:\Theta_{t}\times\mathcal{U}_{t}\rightarrow\Theta_{t+1}\times\mathcal{U}_{t+1}$
is selected from $M_{t}$ possible candidates $\left\{ G_{t\rightarrow t+1}^{(m)}\right\} _{m=1}^{M_{t}}$.
The label of such transformation, denoted by $l_{t}$, is jointly
drawn with $u_{t}$ from the distribution $\psi_{t}\left(\cdot\left|\theta_{t}\right.\right)$.
Therefore, after sampling $\left(u_{t},l_{t}\right)\sim\psi_{t}\left(\cdot\left|\theta_{t}\right.\right)$,
the incremental weight at $t$ in the TSMC algorithm is given by 
\begin{align}
\frac{\tilde{\varphi}_{t+1}}{\tilde{\varphi}_{t\rightarrow t+1}}\left(\vartheta_{t\rightarrow t+1}\right) & =\frac{\tilde{\pi}_{t+1}\left(\theta_{t\rightarrow t+1}\left(\vartheta_{t}\right)\right)\psi_{t+1}\left(\left.u_{t\rightarrow t+1}\left(\vartheta_{t}\right)\right|\theta_{t\rightarrow t+1}\left(\vartheta_{t}\right)\right)}{\tilde{\pi}_{t}\left(\theta_{t}\right)\psi_{t}\left(u_{t},l_{t}\left|\theta_{t}\right.\right)\left|J_{t+1\rightarrow t}^{\left(l_{t}\right)}\right|M_{t}},\label{eq:incWeight_multTransf}
\end{align}
where $J_{t+1\rightarrow t}^{\left(m\right)}$ denotes the Jacobian
of $G_{t+1\rightarrow t}^{(m)}$. Notice that the denominator contains
the term $M_{t}$ since we have introduced the the extra variable
$L_{t}$ in the proposal; thus, in order to obtain the correct ratio
of normalising constants we need to extend the target using a ``dummy''
distribution for $L_{t}$, in this case such distribution is uniform
on the set $\left\{ 1,\dots,M_{t}\right\} $.

The split move from \citet{Richardson1997} clearly falls into this
category since the selected component to be split is chosen uniformly,
i.e. $M_{t}=t$ and 
\[
\psi_{t}\left(u,l\left|\theta_{t}\right.\right)=\frac{1}{t}\psi_{t}^{(s)}\left(u\left|\theta_{t}\right.\right),
\]
for $u\in\mathcal{U}_{t}$ and $l\in\left\{ 1,\dots,t\right\} $;
in this case $\psi_{t}^{(s)}$ is the distribution on the auxiliary
variables $U_{t}$ required for implementing the split move. An improvement
on this idea would be to use a mixture representation of the proposal
as done in Population Monte Carlo \citep{Douc2007a}, i.e. the denominator
of (\ref{eq:incWeight_multTransf}) would become
\begin{equation}
\varphi_{t\rightarrow t+1}\text{\ensuremath{\left(\vartheta_{t\rightarrow t+1}\right)}}=\pi_{t}\left(\theta_{t}\right)\sum_{l=1}^{M_{t}}\psi_{t}\left(u_{t},l\left|\theta_{t}\right.\right)\left|J_{t+1\rightarrow t}^{\left(l\right)}\right|;\label{eq:marginal_denominator}
\end{equation}
however, we do not follow such approach. Instead, we try to alleviate
a possible complication when implementing the split move. After selecting
and splitting the $k$-th component $\left(w_{k},\mu_{k},\tau_{k}\right)$,
two new weights (say $w_{k^{-}}$ and $w_{k^{+}}$), two new means
(say $\mu_{k^{-}}$ and $\mu_{k^{+}}$) and two new precisions (say
$\tau_{k^{-}}$ and $\tau_{k^{+}}$) are obtained. However, if either
\[
\mu_{k^{-}}\notin\left[\mu_{k-1},\mu_{k+1}\right]\qquad\text{or}\qquad\mu_{k^{+}}\notin\left[\mu_{k-1},\mu_{k+1}\right],
\]
then the incremental weight will be zero since the support of the
target $\pi_{t+1}$ has been restricted to ordered means. We solve
this by reordering all the components with respect to their means
and correcting the incremental weight with an extra factor. The correct
incremental weight can be expressed as follows
\begin{align}
\frac{\tilde{\varphi}_{t+1}}{\tilde{\varphi}_{t\rightarrow t+1}} & \left(\vartheta_{t\rightarrow t+1}\right)=\frac{\tilde{\pi}_{t+1}\left(o_{t+1}\left(\theta_{t\rightarrow t+1}\left(\vartheta_{t}\right)\right)\right)\psi_{t+1}\left(\left.u_{t\rightarrow t+1}\left(\vartheta_{t}\right)\right|\theta_{t\rightarrow t+1}\left(\vartheta_{t}\right)\right)}{\tilde{\pi}_{t}\left(\theta_{t}\right)\psi_{t}\left(u_{t},l_{t}\left|\theta_{t}\right.\right)\left|J_{t+1\rightarrow t}^{\left(l_{t}\right)}\right|\left(\begin{array}{c}
t+1\\
2
\end{array}\right)},\label{eq:incWeight_orderedSplit}
\end{align}
where the function $o_{t+1}:\Theta_{t+1}\rightarrow\Theta_{t+1}$
simply combines the two newly created components, $\left(w_{k^{-}},\mu_{k^{-}},\tau_{k^{-}}\right)$
and $\left(w_{k^{+}},\mu_{k^{+}},\tau_{k^{+}}\right)$, with the set
of already ordered $t-1$ components (those that were not split).

To see why (\ref{eq:incWeight_orderedSplit}) is correct we follow
a similar reasoning for deriving (\ref{eq:incWeight_multTransf}).
In order to obtain the correct ratio of normalising constants, we
need to introduce a ``dummy'' distribution in the target. When inverting
the split move with rearrangement, two artificial variables are created
denoting the labels of the newly created components. Since $\mu_{k^{-}}<\mu_{k^{+}}$,
a simple choice for the ``dummy'' distribution is a uniform over
the set 
\[
S_{t+1}=\left\{ \left.\left(h,k\right)\right|h,k\in\left\{ 1,\dots,t+1\right\} \text{ and }h<k\right\} ,
\]
for which $\left|S\right|=\left(\begin{array}{c}
t+1\\
2
\end{array}\right)$, as included in (\ref{eq:incWeight_orderedSplit}).

\subsection{Birth move\label{subsec:Birth-TSMC}}

The birth move can benefit also from a reordering of components. The
correct incremental weight is much simpler than in the split case
since the auxiliary variable $U_{t}\sim\psi_{t}^{(b)}\left(\cdot\left|\theta_{t}\right.\right)$
already represents the new component $\left(w_{*},\mu_{*},\tau_{*}\right)$.
Using the same logic as before, when inverting the birth move with
rearrangement an artificial variable is created which denotes the
place of the most recent generated component. Since this label can
take values in $S_{t+1}=\left\{ 1,\dots,t+1\right\} $, the simplest
choice for the ``dummy'' distribution is a uniform over $S_{t+1}$;
therefore, the expression for the incremental weight in this case
is given by
\begin{align}
\frac{\tilde{\varphi}_{t+1}}{\tilde{\varphi}_{t\rightarrow t+1}}\left(\vartheta_{t\rightarrow t+1}\right) & =\frac{\tilde{\pi}_{t+1}\left(o_{t+1}\left(\theta_{t\rightarrow t+1}\left(\vartheta_{t}\right)\right)\right)\psi_{t+1}\left(\left.u_{t\rightarrow t+1}\left(\vartheta_{t}\right)\right|\theta_{t\rightarrow t+1}\left(\vartheta_{t}\right)\right)}{\tilde{\pi}_{t}\left(\theta_{t}\right)\psi_{t}^{(b)}\left(u_{t}\left|\theta_{t}\right.\right)\left|J_{t+1\rightarrow t}\right|\left(t+1\right)}.\label{eq:incWeight_orderedBirth}
\end{align}

\subsection{Marginalisation of moves\label{subsec:Improvement-marginal}}

The previous descriptions of the birth and split moves are based on
the idea of extending the target using an auxiliary distribution for
the labels created due to the reordering process. We saw that a simple
choice for this auxiliary distributions is a discrete uniform over
the set of possible values for the labels, reason why the weights
in (\ref{eq:incWeight_orderedSplit}) and (\ref{eq:incWeight_orderedBirth})
contain the denominator terms $\left(\begin{array}{c}
t+1\\
2
\end{array}\right)$ and $t+1$, respectively. However, as discussed later in the examples
of Section \ref{subsec:Results}, the corresponding estimators of
the normalising constant may suffer from a very high variance making
them useless from a practical point of view. A way around this problem
is to marginalise the proposal over the artificial label created by
the reordering process; such marginalisation is similar to (\ref{eq:marginal_denominator})
and is now described.

The ordering function $o_{t+1}:\Theta_{t+1}\rightarrow\Theta_{t+1}$,
introduced previously, simply reorders the newly generated component
(or components) from the birth (split) move. In order to be able to
compute the inverse transformation of this reordering, an artificial
variable $\bar{l}_{t+1}\in\bar{S}_{t+1}$ is created which simply
denotes the place (or places) of the new component(s). To be more
precise, there are two transformations applied to $\vartheta_{t}$
that allow us to obtain the final $\vartheta_{t\rightarrow t+1}$
together with the label $\bar{l}_{t+1}$. Let

\[
\bar{G}_{t\rightarrow t+1}\text{\ensuremath{\left(\vartheta_{t}\right)}}:=\bar{o}_{t+1}\circ G_{t\rightarrow t+1}\text{\ensuremath{\left(\vartheta_{t}\right)}}=\bar{o}_{t+1}\left(\theta_{t\rightarrow t+1}\left(\vartheta_{t}\right),u_{t\rightarrow t+1}\left(\vartheta_{t}\right)\right)=\left(o_{t+1}\left(\theta_{t\rightarrow t+1}\left(\vartheta_{t}\right)\right),u_{t\rightarrow t+1}\left(\vartheta_{t}\right),\bar{l}_{t+1}\right)=\left(\vartheta_{t\rightarrow t+1},\bar{l}_{t+1}\right),
\]
where $\bar{o}_{t+1}:\Theta_{t+1}\times\mathcal{U}_{t+1}\rightarrow\Theta_{t+1}\times\mathcal{U}_{t+1}\times\bar{S}_{t+1}$
is an extension of $o_{t+1}$ that reorders $\theta_{t\rightarrow t+1}\left(\vartheta_{t}\right)$,
leaves $u_{t\rightarrow t+1}\left(\vartheta_{t}\right)$ unchanged,
and creates $\bar{l}_{t+1}$. In the previous sections there was no
need to introduce $\bar{l}_{t+1}$ since the denominator in (\ref{eq:incWeight_orderedSplit})
and (\ref{eq:incWeight_orderedBirth}) is obtained simply by transforming
back $\bar{\vartheta}_{t\rightarrow t+1}$ into $\vartheta_{t}$;
observe that for such cases
\begin{align*}
\varphi_{t\rightarrow t+1}\text{\ensuremath{\left(\vartheta_{t\rightarrow t+1}\right)}}= & \varphi_{t}\left(\bar{G}_{t+1\rightarrow t}\left(\vartheta_{t\rightarrow t+1},\bar{l}_{t+1}\right)\right)\left|J_{t+1\rightarrow t}\right|=\varphi_{t}\left(\vartheta_{t}\right)\left|J_{t+1\rightarrow t}\right|.
\end{align*}
The marginalisation step becomes clear by integrating out the variable
$\bar{l}_{t+1}$; in this case the denominators in (\ref{eq:incWeight_orderedSplit})
and (\ref{eq:incWeight_orderedBirth}) respectively become

\begin{align*}
\varphi_{t\rightarrow t+1}\text{\ensuremath{\left(\vartheta_{t\rightarrow t+1}\right)}}= & \sum_{\bar{l}_{t+1}\in\bar{S}_{t+1}}\varphi_{t}\left(\bar{G}_{t+1\rightarrow t}\left(\vartheta_{t\rightarrow t+1},\bar{l}_{t+1}\right)\right)\left|J_{t+1\rightarrow t}^{(l_{t})}\right|=\sum_{\bar{l}_{t+1}\in\bar{S}_{t+1}}\pi_{t}\left(\theta_{t}\right)\bar{\psi}_{t}\left(u_{t},l_{t}\left|\theta_{t}\right.\right)\left|J_{t+1\rightarrow t}^{(l_{t})}\right|\\
\text{and}\qquad\varphi_{t\rightarrow t+1}\text{\ensuremath{\left(\vartheta_{t\rightarrow t+1}\right)}}= & \sum_{\bar{l}_{t+1}\in\bar{S}_{t+1}}\varphi_{t}\left(\bar{G}_{t+1\rightarrow t}\left(\vartheta_{t\rightarrow t+1},\bar{l}_{t+1}\right)\right)\left|J_{t+1\rightarrow t}\right|=\sum_{\bar{l}_{t+1}\in\bar{S}_{t+1}}\pi_{t}\left(\theta_{t}\right)\psi_{t}^{(b)}\left(u_{t}\left|\theta_{t}\right.\right)\left|J_{t+1\rightarrow t}\right|,
\end{align*}
recalling that the variables $\theta_{t}$, $u_{t}$ and $l_{t}$
depend on $\left(\vartheta_{t\rightarrow t+1},\bar{l}_{t+1}\right)$
via the inverse transformation $\bar{G}_{t+1\rightarrow t}$.

In Section \ref{subsec:Results}, we look at the performance of the
marginalised versions of the birth and split moves against those described
in Sections \ref{subsec:Split-TSMC} and \ref{subsec:Birth-TSMC},
which we term the conditional versions. It is clear that marginalising
should be a sensible approach for reducing the variance of the estimated
of the normalising constants, however in certain cases obtaining such
marginal could become expensive or impractical if the required sum
contains a large number of elements.

\subsection{Details on the MCMC moves\label{subsec:MCMC-details}}

The MCMC moves are performed in the transformed space of logit-weights,
means and log-precisions. Given a set of $t$ components $\left\{ \left(w_{k},\mu_{k},\tau_{k}\right)\right\} _{k=1}^{t}$
at time $t$, the transformed components $\left\{ \left(lw_{k},\mu_{k},l\tau_{k}\right)\right\} _{k=1}^{t}$
are given by 
\begin{align*}
lw_{k} & :=\log\left(\frac{w_{k}}{w_{t}}\right)=\log\left(\frac{w_{k}}{1-\sum_{j=1}^{t-1}w_{j}}\right),\\
l\tau_{k} & :=\log\left(\tau_{k}\right).
\end{align*}

We consider two scenarios. For the first one we implement an adaptive
Gaussian random-walk Metropolis algorithm on the transformed space,
taking into account the Jacobian of the previous transformation. The
adaptation is done in the proposal variance-covariance matrix in such
way that the estimated acceptance probability from the particles stays
near 0.20. More precisely, an initial diagonal variance-covariance
matrix for the logit-weights, means and log-precisions is selected
(say $\Sigma_{prop}^{(0)}$); then, after propagating the $N$ particles,
an estimated acceptance probability is obtained (say $\hat{\alpha}_{P}^{(0)}$).
If such estimation lies outside a neighbourhood of 0.20, then a new
matrix is obtained as follows
\begin{align*}
\Sigma_{prop}^{(1)} & =\frac{\hat{\alpha}_{P}^{(0)}}{0.20}\Sigma_{prop}^{(0)}.
\end{align*}
The process starts again (and is repeated until the desired acceptance
probability is achieved) by propagating particles using $\Sigma_{prop}^{(1)}$
and computing the estimated acceptance $\hat{\alpha}_{P}^{(1)}$.
One should be careful not to take a small number of particles or a
small neighbourhood around 0.20 since the number of adaptations needed
may be large. As seen in the following section, the previous choice
of proposal could be quite inefficient since the particles may not
move far from their current value. Nevertheless, using such an inefficient
proposal will allow us to empirically quantify the effects of good
and bad transformations $G_{t\rightarrow t+1}$: we present the results
using this proposal in the following section.

For the second scenario (the one used in the results in the main body
of the paper), using the set of particles approximately distributed
according to $\varphi_{t\rightarrow t+1,k}$ we compute the empirical
variance-covariance matrix $\hat{\Sigma}_{t+1,k}$. The proposal variance-covariance
matrix for targeting $\varphi_{t\rightarrow t+1,k+1}$ is then chosen
as follows $\Sigma_{prop}=\hat{\Sigma}_{t+1,k}/(t+1)$ . This choice
will certainly be a more sensible proposal provided a Gaussian proposal
is able to capture the shape of the target and the consecutive intermediate
distributions are similar; as seen in the following section, a carefully
designed MCMC kernel (together with a good transformation $G_{t\rightarrow t+1}$)
can dramatically improve the quality of the particles.

\subsection{Results\label{subsec:Results}}

This section shows results from SMC2 and the TSMC algorithms on the
enzyme, acidity and galaxy data from \citet{Richardson1997} (see
figure xxx). We ran the algorithms 50 times, up to a maximum of $8$
components, with 500 particles. We used an adaptive sequence of intermediate
distributions, choosing the next intermediate distribution to be the
one the yields a CESS of $\beta P$, where $\beta=0.99$. We resampled
using stratified resampling when the ESS falls below $\alpha P$,
where $\alpha=0.5$. The first adaptive MCMC scheme was used.

\begin{figure}
\includegraphics[scale=0.13]{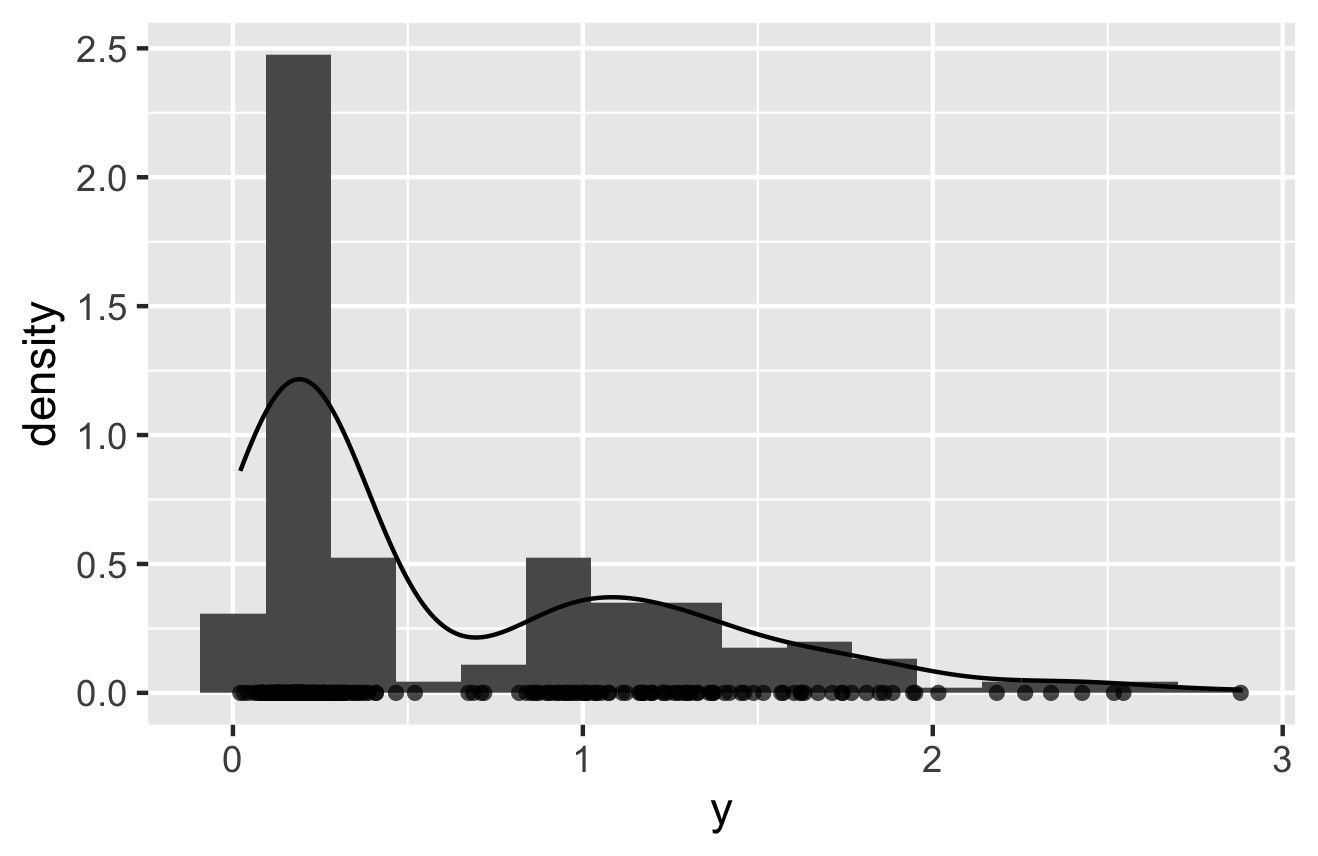}\includegraphics[scale=0.13]{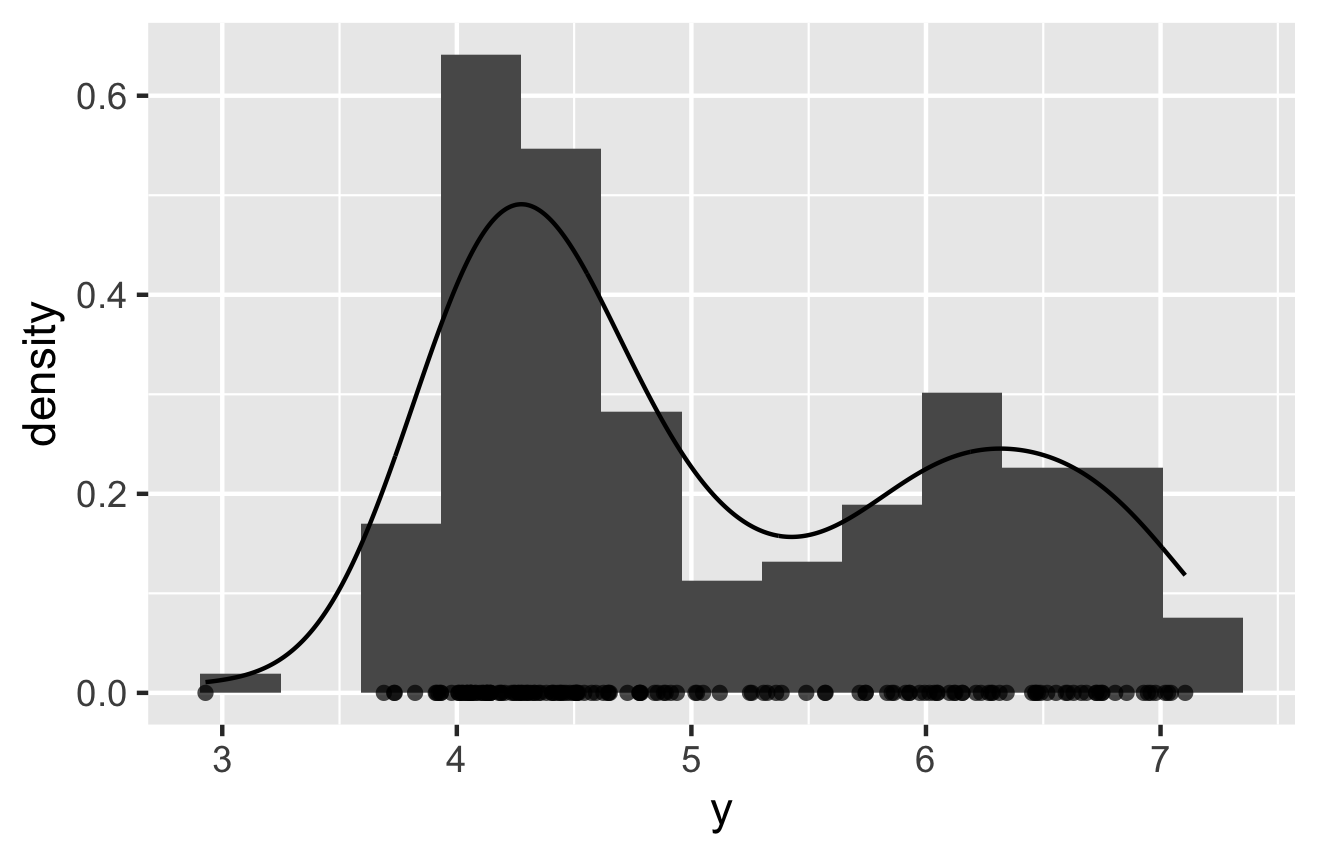}\includegraphics[scale=0.13]{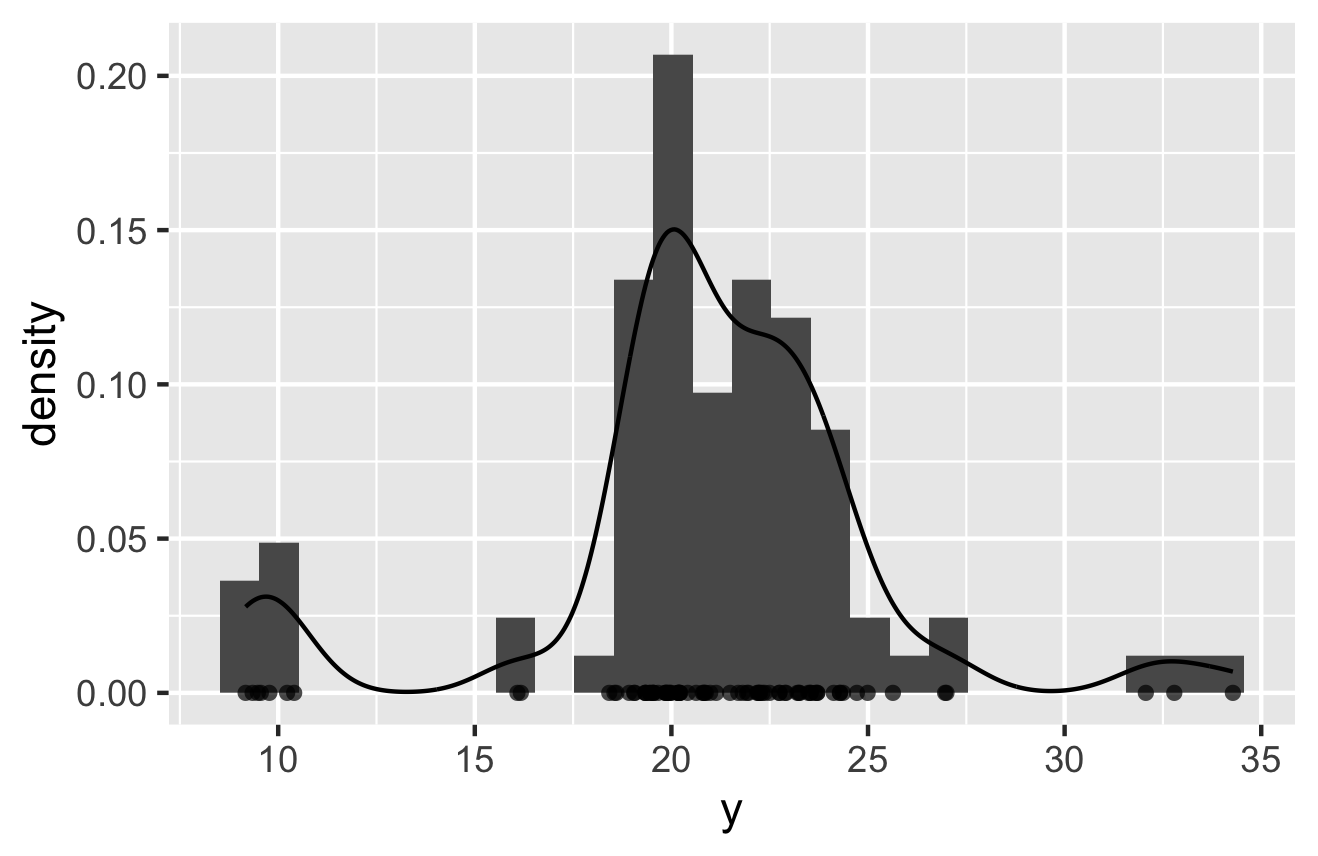}

\caption{Density plots of the enzyme, acidity and galaxy data.}

\end{figure}

\begin{figure*}

\begin{tabular}{cc}

\begin{subfigure}[t]{0.5\textwidth}
        \centering
        \includegraphics[width=1.2\textwidth]{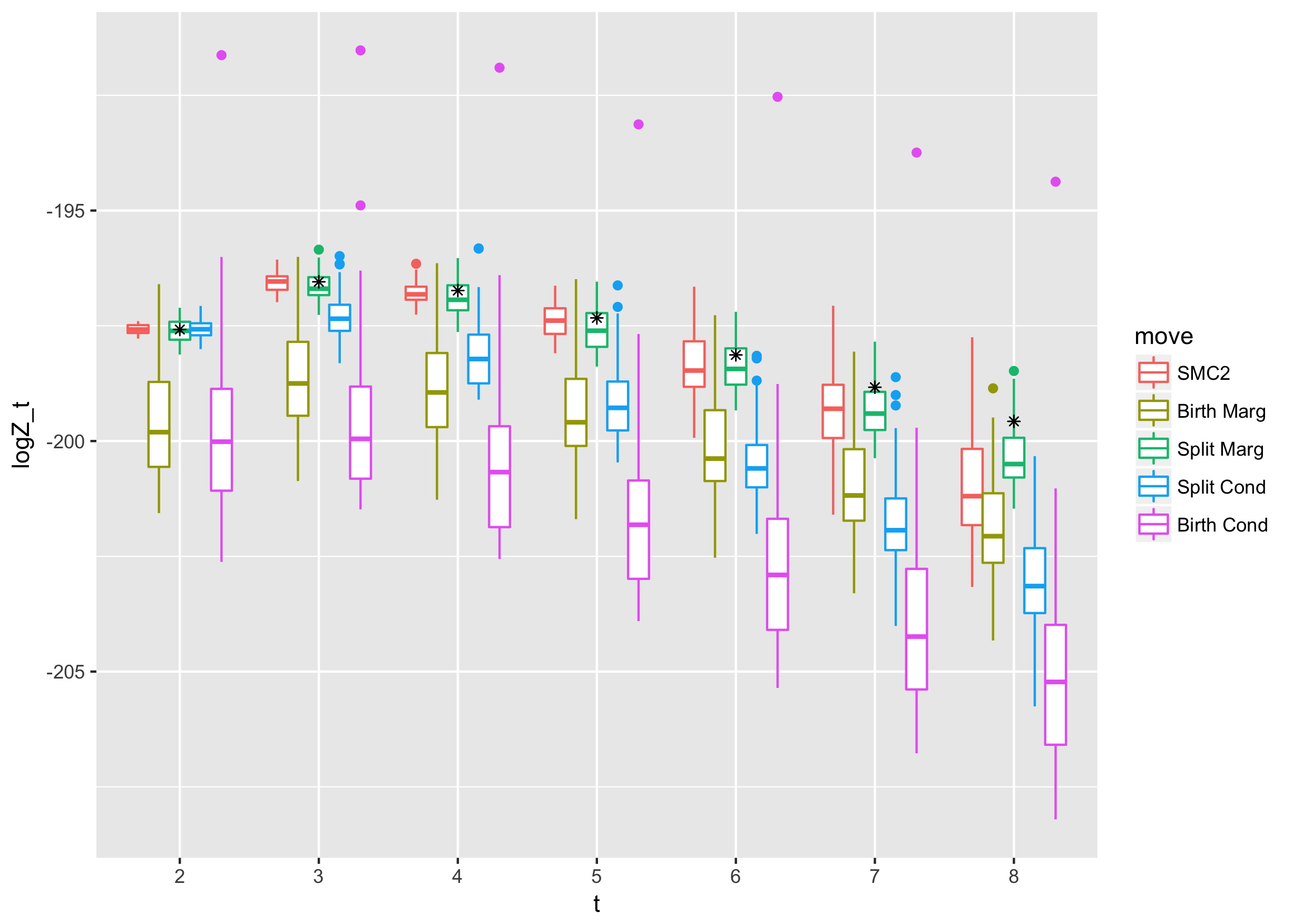}
        \caption{Box plots of the log marginal likelihood estimates from each algorithm.
Black dots represent the ``truth'' computed using a long SMC2 run.}
        \label{fig:Box-plots-of-1}
\end{subfigure}

\begin{subfigure}[t]{0.5\textwidth}
        \centering
        \includegraphics[width=1.2\textwidth]{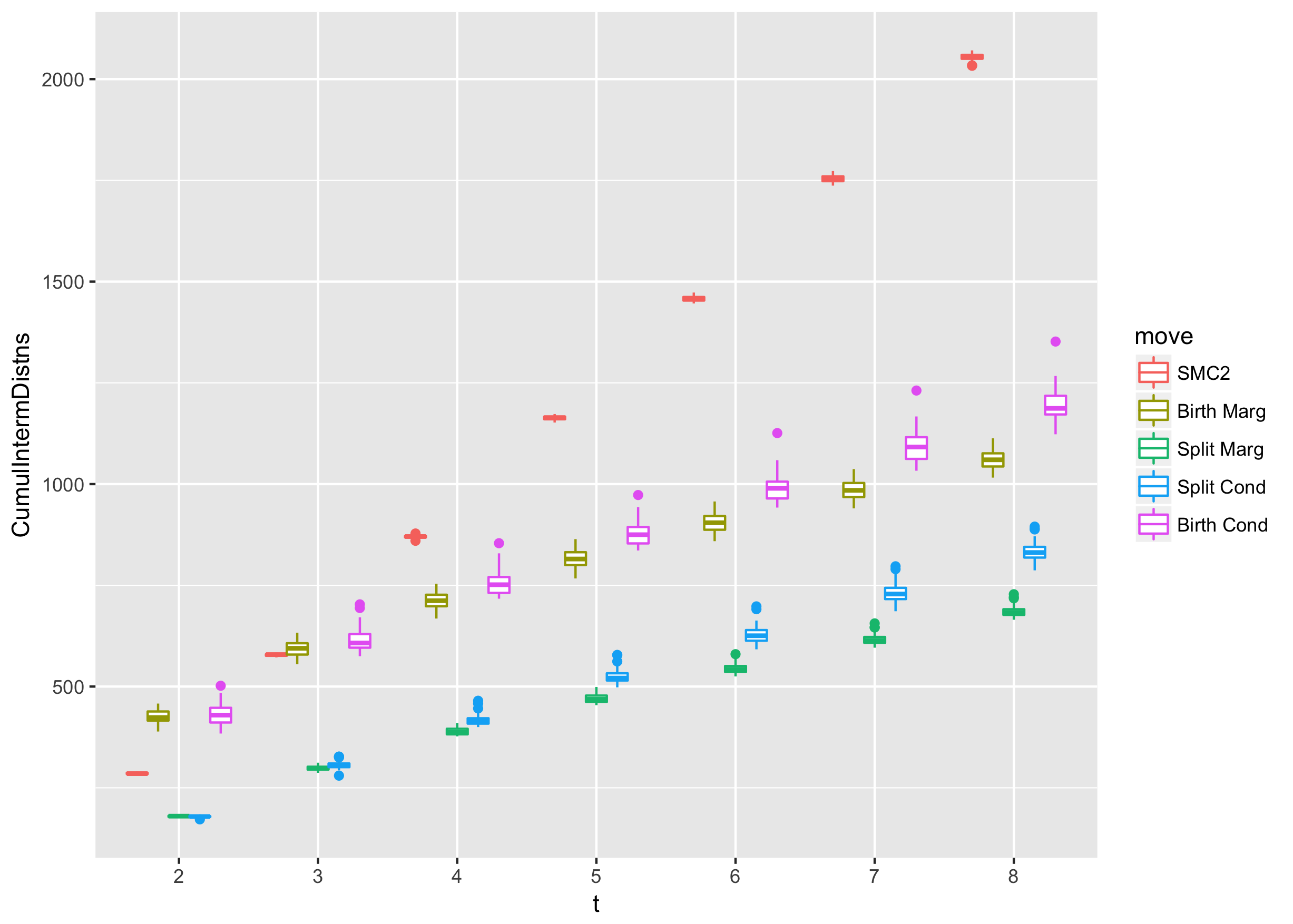}
        \caption{The cumulative number of intermediate distributions up to model t.}
        \label{fig:The-number-of-1}
\end{subfigure}

\end{tabular}
\caption{The relative performance of the different SMC schemes on the acidity
data.}
\label{fig:The-relative-performance-1}
\end{figure*}

\begin{figure*}

\begin{tabular}{cc}

\begin{subfigure}[t]{0.5\textwidth}
        \centering
        \includegraphics[width=1.2\textwidth]{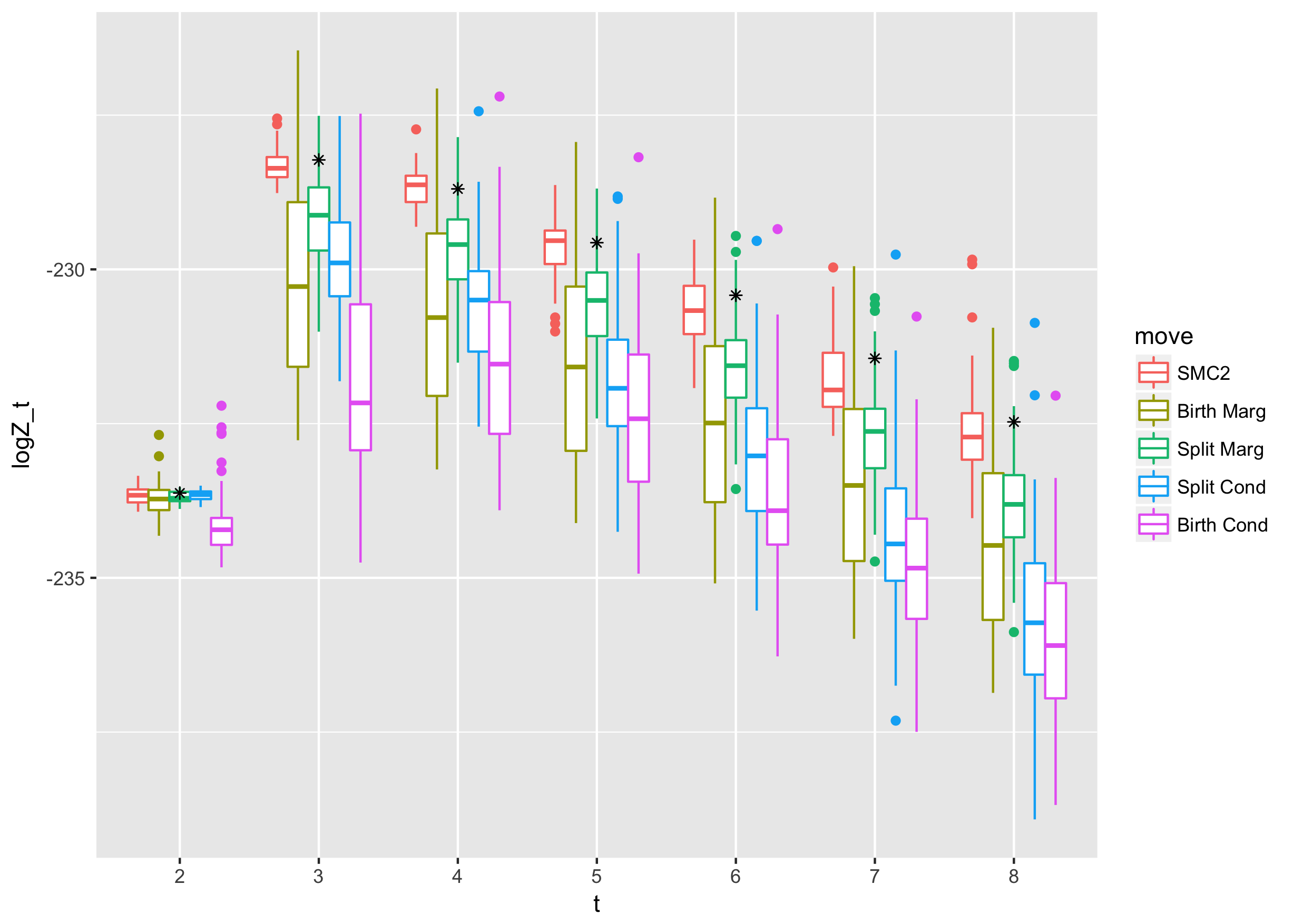}
        \caption{Box plots of the log marginal likelihood estimates from each algorithm.
Black dots represent the ``truth'' computed using a long SMC2 run.}
        \label{fig:Box-plots-of-2}
\end{subfigure}

\begin{subfigure}[t]{0.5\textwidth}
        \centering
        \includegraphics[width=1.2\textwidth]{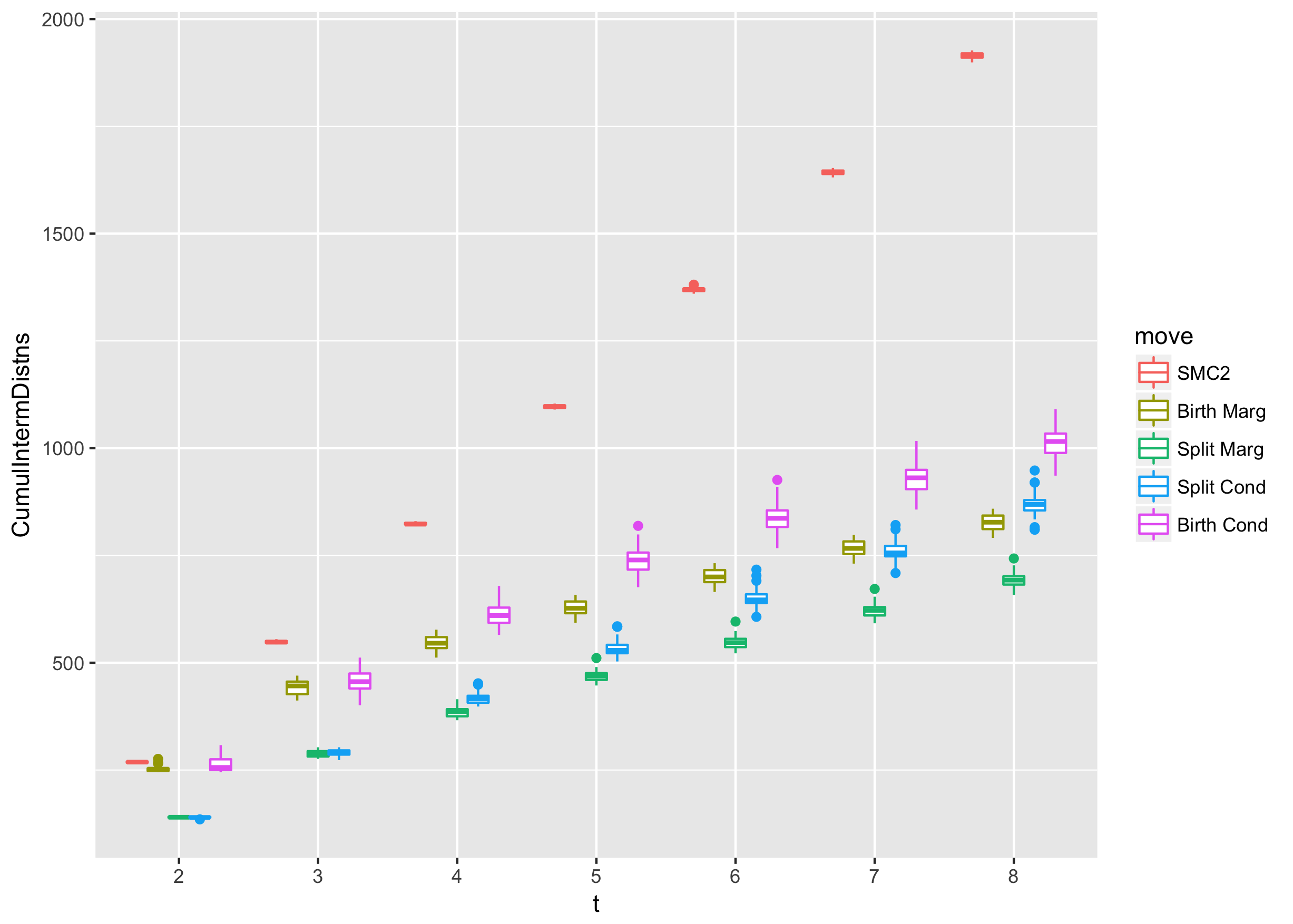}
        \caption{The cumulative number of intermediate distributions up to model t.}
        \label{fig:The-number-of-2}
\end{subfigure}

\end{tabular}
\caption{The relative performance of the different SMC schemes on the galaxy
data.}
\label{fig:The-relative-performance-2}
\end{figure*}

Figures \ref{fig:The-relative-performance}, \ref{fig:The-relative-performance-1}
and \ref{fig:The-relative-performance-2} show log marginal likelihood
estimates from the different approaches, and the cumulative number
of intermediate distributions used in estimating all of the marginal
likelihoods up to model $k$ for each $k$. The key observations from
these results are:
\begin{itemize}
\item The less effective adaptive MCMC scheme has a negative impact on the
results (comparing figure \ref{fig:The-relative-performance} with
those the in the main paper). Despite this, the marginal split TSMC
still exhibits good performance on the enzyme and acidity data sets
(in contrast to SMC2).
\item TSMC appears to be less effective, compared to SMC2, on the galaxy
data. When using the birth move, the reason is the same as stated
in the main body of the paper: that the posterior on the parameters
of existing components in model $t$ does not provide a good proposal
for these parameters in model $t+1$. For the split move, the results
can be explained by the distribution of the data. This data set contains
a few points located at a relatively large distance from the rest
of the points in the dataset. The higher order models place components
on these small clusters of points, whilst maintaining the most important
components in the centre. In this case the split move is not an effective
way to move between distributions (as also observed in \citet{Richardson1997}),
thus the performance of TSMC with a split move is not as effective
as SMC2, which uses the prior as the proposal for the parameters of
all components.
\end{itemize}

\begin{figure*}

\begin{tabular}{cc}

\begin{subfigure}[t]{0.5\textwidth}
        \centering
        \includegraphics[width=1.2\textwidth]{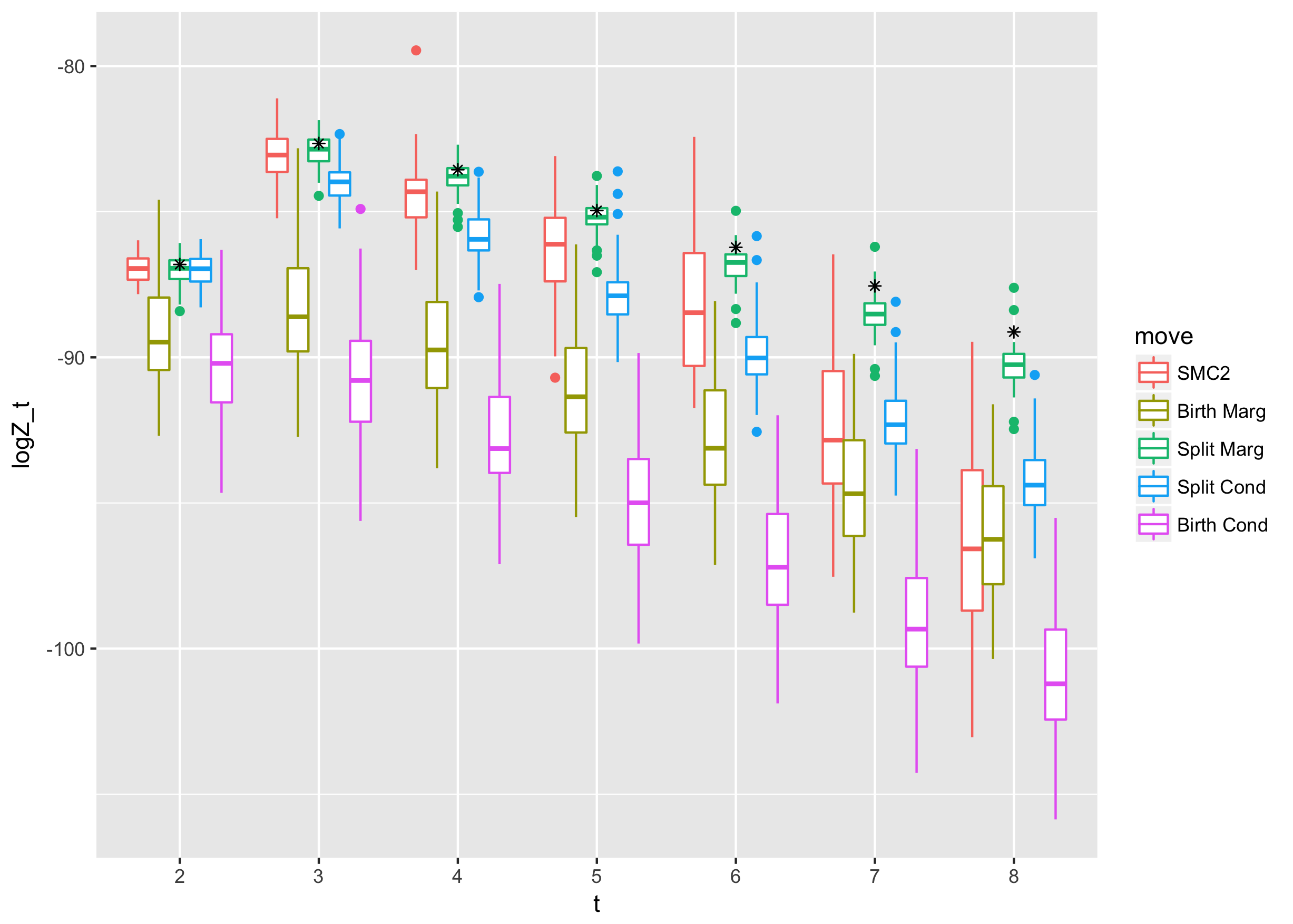}
        \caption{Box plots of the log marginal likelihood estimates from each algorithm.
Black dots represent the ``truth'' computed using a long SMC2 run.}
        \label{fig:Box-plots-of}
\end{subfigure}

\begin{subfigure}[t]{0.5\textwidth}
        \centering
        \includegraphics[width=1.2\textwidth]{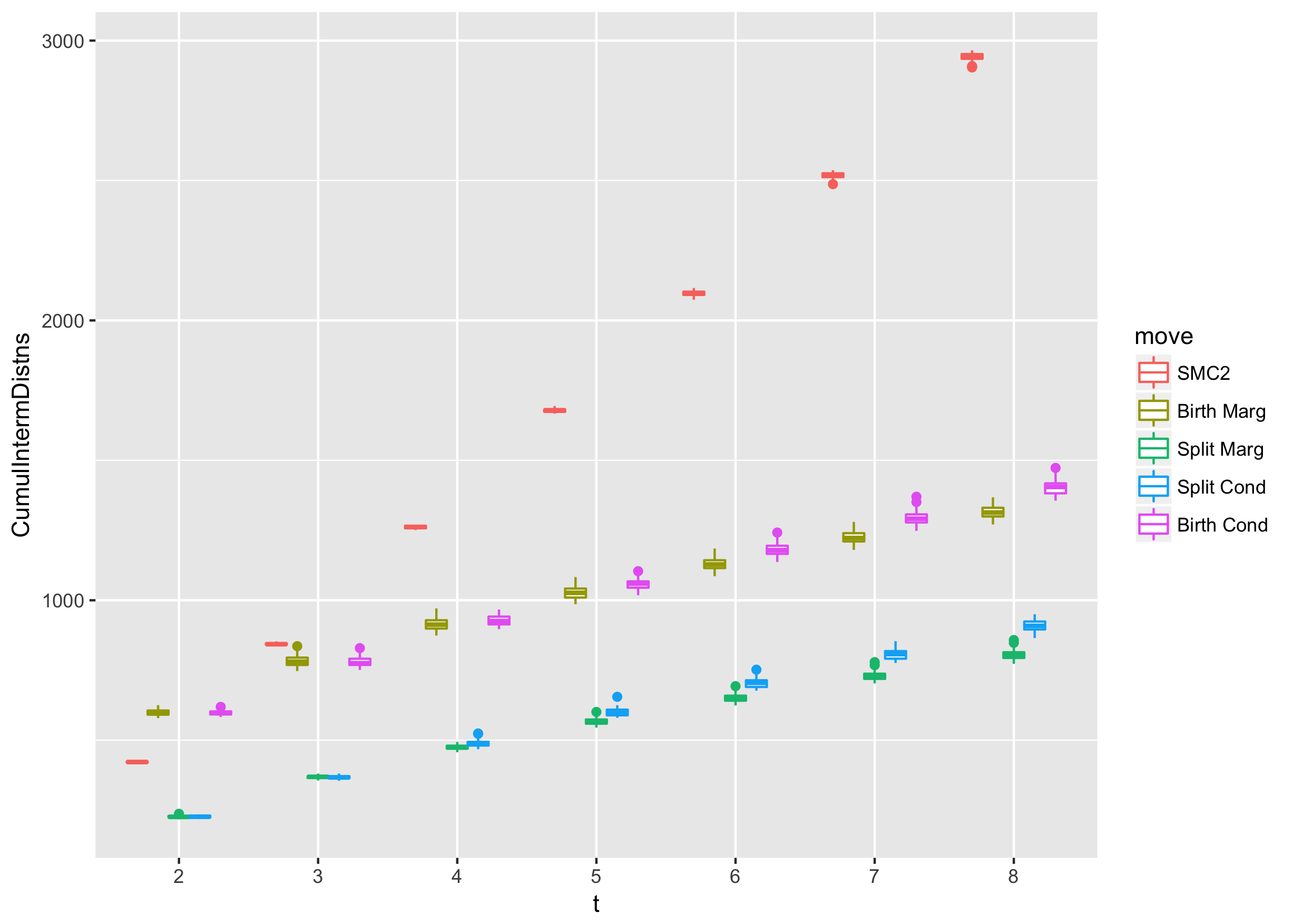}
        \caption{The cumulative number of intermediate distributions up to model t.}
        \label{fig:The-number-of}
\end{subfigure}

\end{tabular}
\caption{The relative performance of the different SMC schemes on the enzyme
data.}
\label{fig:The-relative-performance}
\end{figure*}

\section{Sequential Bayesian inference under the coalescent\label{sec:Sequential-Bayesian-inference}}

\subsection{Transformation and weight update}

Let $g_{\text{t}}\sim\chi_{t}^{(g)}\left(\cdot\mid\theta_{t},\mathcal{T}_{t},y_{1:t+1}\right)$
and $h_{t}^{(\text{new})}\sim\chi_{t}^{(h)}\left(\cdot\mid g_{t},\theta_{t},\mathcal{T}_{t},,y_{1:t+1}\right)$.
The transformation $G_{t\rightarrow t+1}$ leaves $\theta$, and $g_{s},h_{s}^{(\text{new})}$
for $s>t$, unchanged. It makes a new tree from $\left(\mathcal{T}_{t},g_{t},h_{t}^{(\text{new})}\right)$
as follows. Firstly, $g_{t}$ chooses a lineage to add the new branch
to, where each possible lineage is indexed by the leaf on that lineage.
Next we examine the coalescent heights. If $\iota$ is such that $h_{t}^{(\iota+1)}<h_{t}^{(\text{new})}<h_{t}^{(\iota)}$
then the effect of the transformation on the coalescence heights is
\[
\left(\left(h_{t}^{(t)},...,h_{t}^{(2)}\right),h_{t}^{(\text{new})}\right)\mapsto\left(h_{t}^{(t)},...,h_{t}^{(\iota+1)},h_{t}^{(\text{new})},h_{t}^{(\iota)},...,h_{t}^{(2)}\right)
\]
(or adding the new height to the beginning or the end of the vector
if it is the first or last coalescence event), giving a Jacobian of
1. Then the new branching order is given by the original branching
order, with a split in the branch that is uniquely determined by $\left(\iota,g_{t}\right)$,
where the new branching order variable is denoted $b_{t}^{(\text{new})}$.
We note that this transformation is not bijective: since branches
higher up the tree are shared by multiple lineages there are multiple
possible lineages that could have led to each tree.

Without loss of generality we examine the case of no intermediate
distributions. As noted in the paper, $g_{s},h_{s}^{(\text{new})}$
for $s>t$ are not involved in the weight update. The variables involved
are $\mathcal{T}_{t+1},\theta$, which have resulted from the application
of $G_{t\rightarrow t+1}$. To find $\varphi_{t\rightarrow t+1}$
we must find the distribution under $\varphi_{t}$ of the inverse
image of $\mathcal{T}_{t+1},\theta$. The resultant weight update
is
\begin{equation}
\tilde{w}_{t+1}=w_{t}\frac{\pi_{t+1}\left(\mathcal{T}_{t+1},\theta\mid y_{1:t+1}\right)}{\pi_{t}\left(\mathcal{T}_{t},\theta\mid y_{1:t}\right)\left[\sum_{s\in\Lambda}\chi_{t}^{(g)}\left(g_{t}=s\mid\theta_{t},\mathcal{T}_{t},y_{1:t+1}\right)\chi_{t}^{(h)}\left(h_{t}^{(\text{new})}\mid g_{t}=s,\theta_{t},\mathcal{T}_{t},y_{1:t+1}\right)\right]},\label{eq:lineage_then_height-weight}
\end{equation}
where $\Lambda$ is the set that contains the leaves of the lineages
that could have resulted in $b_{t}^{(\text{new})}$. Note the relationship
with the Rao-Blackwellised weight update described in the paper: we
achieve a lower variance through summing over the possible lineages
rather than using an SMC over the joint space that includes the lineage
variable.

\subsection{Design of auxiliary distributions}

For our SMC sampler to be efficient, we must design $\chi_{t}^{(g)}$
and $\chi_{t}^{(h)}$ such that the distributions in the numerator
and denominator of (\ref{eq:lineage_then_height-weight}) are close,
i.e. resulting in many trees that have high probability under the
posterior with $t+1$ sequences, but with the denominator having heavier
tails than the numerator.

To choose the lineage, we make use of an approximation to the probability
that the new sequence is $M_{s}$ mutations from each of the existing
leaves. Following \citet{Stephens2000c} (also see \citealp{Li2003})
we choose the probability of choosing the lineage with leaf $s$ using
\begin{equation}
\chi_{t}^{(g)}\left(s\mid\theta_{t},y_{1:t+1}\right)\propto\left(\frac{N\theta_{t}}{t+N\theta_{t}}\right)^{M_{s}}.\label{eq:prob_the_same_coal}
\end{equation}
This probability results from using a geometric distribution on the
number of SNP differences between the new sequence and sequence $s$
for each $s$, which is a generalisation of Ewens' sampling formula
\citep{Ewens1972} to the finite allele case. The geometric distribution
results from integrating over possible coalescence times of the new
sequence (where distribution on the time is modelled as exponentially
distributed with the correct mean), yielding a choice for $\chi_{t}^{(g)}$
that is likely to give our importance sampling proposal a larger variance
than our target.

For $\chi_{t}^{(h)}$ we propose to approximate the pairwise likelihood
$f_{t+1,s}\left(y_{s},y_{t+1}\mid\theta,h_{t}^{(\text{new})},g_{t}=s\right)$,
where $y_{s}$ is the sequence at the leaf of the chosen lineage.
Since only two sequences are involved in this likelihood, it is likely
to have heavier tails than the posterior. Our pairwise likelihood
is
\[
L\left(h_{\text{t}}^{(new)}\mid y_{s},y_{t+1},\theta_{t}\right)=\left[\frac{3}{4}-\frac{3}{4}\exp\left(-4\theta_{t}h_{\text{t}}^{(\text{new})}/3\right)\right]^{M_{s}}\left[\frac{1}{4}+\frac{3}{4}\exp\left(-4\theta_{t}h_{\text{t}}^{(\text{new})}/3\right)\right]^{N-M_{s}},
\]
where $M_{s}$ is the number of pairwise SNP differences between the
new sequence and sequence $s$, both of length $N$. This likelihood
may be approximated by a distribution using the Laplace approximation
$\mathcal{N}\left(\mu=\hat{h},\sigma^{2}=\left(-\hat{H}\right)^{-1}\right)$,
where $\hat{h}_{t}^{(new)}$ denotes the maximum likelihood estimate
of $h_{t}^{(\text{new})}$ and $\hat{H}$ an estimate of the Hessian
of the log likelihood at this estimate \citep{Bishop2006c}. \citet{Reis2011}
proposes an accurate approximation of the two sequence likelihood
by using a Laplace approximation in a transformed space, in particular
they propose to use the transformation $2\arcsin\sqrt{\frac{3}{4}-\frac{3}{4}\exp\left(-2\theta_{t}h_{t}^{(\text{new})}/3\right)}$.
In this case the mean and variance of the Gaussian approximation are
respectively $\mu=\hat{h}$ and $\sigma^{2}=\left(-\hat{H}\right)^{-1}$
where $\hat{h}=2\arcsin\sqrt{\frac{M_{s}}{N}}$ and $\hat{H}=-N$.
Thus in order to simulate a new height $h^{(\text{new})}$ we first
simulate $\beta\sim\mathcal{N}\left(2\arcsin\sqrt{\frac{M_{s}}{N}},1/N\right)$
and then compute
\begin{equation}
h_{t}^{(\text{new})}=-\frac{3}{4\theta_{t}}\log\left(1-\frac{4}{3}\sin^{2}\left(\beta/2\right)\right).
\end{equation}
 The density of this distribution is given by
\begin{eqnarray*}
\chi_{t}^{(h)}\left(h_{t}^{(\text{new})}\right) & = & \left|\frac{2\theta\exp\left\{ -\frac{4}{3}\theta h_{t}^{(\text{new})}\right\} }{\sqrt{3}\sqrt{1-\frac{3\left(1-\exp\left\{ -\frac{4}{3}\theta h_{t}^{(\text{new})}\right\} \right)}{4}}\sqrt{1-\exp\left\{ -\frac{4}{3}\theta h_{t}^{(\text{new})}\right\} }}\right|\\
 &  & \times\frac{1}{\sigma\sqrt{2\pi}}\exp\left(-\frac{\left(2\arcsin\sqrt{\frac{3}{4}\left(1-\exp\left\{ -\frac{4}{3}\theta h_{t}^{(\text{new})}\right\} \right)}-\mu\right)^{2}}{2\sigma^{2}}\right).
\end{eqnarray*}

\subsection{SMC and MCMC details\label{subsec:SMC-and-MCMC}}

Our MCMC moves when moving from target $t$ to $t+1$ are technically
made on the space $E_{t+1}$, and in practice made on the space $\Theta_{t+1}$
(recalling that the variables $u_{t+1\rightarrow t}$ will be updated
by direct simulation from $\psi_{t+1}$). The default configuration
of our method was as follows. We use the following moves on each parameter
in $\Theta_{t+1}$: for $\theta$ we use a multiplicative random walk,
i.e. an additive normal random walk in $\log$-space, with proposal
variance $\sigma_{\theta}^{2}$ in $\log$-space; for each height
$h^{(a)}$ ($2<a<T+1$) we use a truncated normal proposal with mean
the current value of $h^{(a)}$ and variance $\sigma_{h^{(a)}}^{2}$.
For the branching order we use 20 subtree pruning and regrafting (SPR)
moves in each sweep of the MCMC: pilot runs suggested that this many
proposed moves result in approximately 0.5 moves being accepted at
each sweep of the MCMC; adaptive methods may also be used to make
such a choice automatically \citep{South2019}.

The SMC uses $P=250$ particles, with an adaptive sequence of intermediate
distributions, choosing the next intermediate distribution to be the
one the yields a CESS of $\beta P$, where $\beta=0.95$. We used
stratified resampling when the ESS falls below $\alpha P$, where
$\alpha=0.5$. At each iteration we used the current population of
particles to tune the proposal variances $\sigma_{\theta}^{2}$ and
$\sigma_{h^{(a)}}^{2}$ for each $a$. Each variance was decomposed
into two terms as follows $\sigma^{2}=s\hat{\sigma^{2}}$, with $\hat{\sigma^{2}}$
being an empirical variance and $s$ being a scaling factor (different
for each proposal variance). $\hat{\sigma_{\theta}^{2}}$ was taken
to be the empirical variance of the weighted particles for $\theta$.
$\hat{\sigma_{h^{(a)}}^{2}}$ was taken to be the empirical variance
of the residuals after using a linear regression of $h^{(a)}$ on
$\theta$ (used due to the strong dependence of $h^{(a)}$ on $\theta$).
Each scaling $s$ was initialised to 1, and: doubled at each iteration
where the acceptance rate was estimated as greater than 0.6; halved
where the rate was estimated at less than 0.15.

\bibliographystyle{spbasic}      
\bibliography{extracted}   

%
%

\end{document}